\documentclass[10pt,journal,english,twocolumn]{IEEEtran}
\usepackage{babel}
\usepackage[babel]{microtype}
\usepackage[babel]{csquotes}

\usepackage[math]{blindtext}

\usepackage[T1]{fontenc}

\usepackage[svgnames]{xcolor}
\usepackage{graphicx}
\usepackage{tikz}
\usepackage{pgfplots}

\usepackage{tabularx}
\usepackage{booktabs}

\usepackage{stfloats}
\usepackage{flushend}

\usepackage{subfigure}

\usepackage{amsmath}
\usepackage{amsfonts}
\usepackage{amssymb}
\usepackage{amsthm}
\usepackage{bm}
\usepackage{siunitx}
\usepackage{mathtools}
\usepackage{dsfont}

\usepackage{algorithm}
\usepackage{algpseudocode}

\usepackage{titlesec}

\usepackage[backend=biber,url=false,style=ieee,bibstyle=ieee-proc,isbn=false,doi=true,eprint=true]{biblatex}
\bibliography{modernize-literature.bib}
\renewcommand*{\bibfont}{\small}

\usepackage[pdfusetitle]{hyperref}
\hypersetup{
	pdfauthor={Karl-Ludwig Besser, Eduard A. Jorswieck, Justin P. Coon},
	pdftitle={An Efficient Frequency Diversity Scheme for Ultra-Reliable Communications in Two-Path Fading Channels},
	colorlinks=true,
	allcolors=.,
	urlcolor=MidnightBlue,
	bookmarksnumbered=true,
	bookmarksopen,
}
\usepackage{bookmark}%
\bookmarksetup{%
	open,%
	openlevel=2,%
	addtohook={%
		\ifnum\bookmarkget{level}=1%
		\bookmarksetup{bold}%
		\fi%
	},%
}

\usepackage[acronyms,nomain,xindy,nowarn]{glossaries}
\makeglossaries
\newacronym{5g}{5G}{Fifth-Generation Wireless Systems}
\newacronym{ask}{ASK}{amplitude-shift keying}
\newacronym[plural={AWGN}, \glsshortpluralkey={AWGN}]{awgn}{AWGN}{additive white Gaussian noise}

\newacronym{ber}{BER}{bit error rate}
\newacronym{bler}{BLER}{block error rate}
\newacronym{bpsk}{BPSK}{binary phase-shift keying}

\newacronym{cdf}{CDF}{cumulative distribution function}
\newacronym{cnn}{CNN}{convolutional neural network}
\newacronym{csi}{CSI}{channel state information}
\newacronym{csit}{CSI\babelhyphen{nobreak}T}{channel-state information at the transmitter}
\newacronym{csir}{CSI\babelhyphen{nobreak}R}{channel-state information at the receiver}

\newacronym{ecc}{ECC}{error-correcting code}
\newacronym{ecdf}{ECDF}{empirical distribution function}
\newacronym{ee}{EE}{energy efficiency}

\newacronym{fdd}{FDD}{frequency-division duplex}
\newacronym{ff}{FF}{feed-forward}
\newacronym{ffnn}{FF-NN}{feed-forward neural network}
\newacronym{fsk}{FSK}{frequency-shift keying}

\newacronym{gm}{GM}{Gaussian mixture}

\newacronym{iid}{i.i.d.}{independent and identically distributed}
\newacronym{il}{IL}{information leakage}
\newacronym{iot}{IoT}{internet of things}

\newacronym{jcas}{JCAS}{joint communications and sensing}

\newacronym{lb}{LB}{lower bound}
\newacronym{los}{LoS}{line-of-sight}

\newacronym{mac}{MAC}{multiple access channel}
\newacronym{map}{MAP}{maximum a posteriori}
\newacronym{mc}{MC}{Monte Carlo}
\newacronym{mimo}{MIMO}{multiple-input multiple-output}
\newacronym{ml}{ML}{machine learning}
\newacronym{mmwave}{mmWave}{millimeter wave}
\newacronym{mop}{MOP}{multi-objective programming problem}
\newacronym{mrc}{MRC}{maximum ratio combining}
\newacronym{msc}{MSC}{modulation and coding scheme}
\newacronym{mse}{MSE}{mean squared error}

\newacronym{nlos}{NLoS}{non-line-of-sight}
\newacronym{nn}{NN}{neural network}
\newacronym{nve}{NVE}{normalized validation error}

\newacronym{ofdm}{OFDM}{orthogonal frequency division multiplexing}

\newacronym{pdf}{PDF}{probability density function}
\newacronym{pwc}{PWC}{polar wiretap code}

\newacronym{qam}{QAM}{quadrature amplitude modulation}

\newacronym{ra}{RA}{rearrangement algorithm}
\newacronym{rbf}{RBF}{radial basis function}
\newacronym{relu}{ReLU}{rectified linear unit}
\newacronym{ris}{RIS}{reconfigurable intelligent surface}
\newacronym{rnn}{RNN}{recurrent neural network}

\newacronym{sde}{SDE}{stochastic differential equation}
\newacronym{sgd}{SGD}{stochastic gradient descent}
\newacronym{sgp}{SGP}{secure goodput}
\newacronym{simo}{SIMO}{single-input multiple-output}
\newacronym{sinr}{SINR}{signal-to-interference-plus-noise ratio}
\newacronym{siso}{SISO}{single-input single-output}
\newacronym{snr}{SNR}{signal-to-noise ratio}
\newacronym{svm}{SVM}{support vector machine}

\newacronym{tvd}{TVD}{total variation distance}

\newacronym{uav}{UAV}{unmanned aerial vehicle}
\newacronym{ub}{UB}{upper bound}
\newacronym{urllc}{URLLC}{ultra-reliable low latency communication}

\newacronym{v2v}{V2V}{vehicle-to-vehicle}

\newacronym{zoc}{ZOC}{zero-outage capacity}
\setacronymstyle{long-short}
\glsdisablehyper

\definecolor{accent}{HTML}{339933}
\definecolor{change}{HTML}{0096b8}

\definecolor{plot0}{HTML}{004488}
\definecolor{plot1}{HTML}{DDAA33}
\definecolor{plot2}{HTML}{BB5566}
\definecolor{plot3}{HTML}{000000}
\definecolor{plot4}{HTML}{AAAAAA}

\definecolor{plot0}{HTML}{EE7733}
\definecolor{plot1}{HTML}{0077BB}
\definecolor{plot2}{HTML}{33BBEE}
\definecolor{plot3}{HTML}{EE3377}
\definecolor{plot4}{HTML}{CC3311}
\definecolor{plot5}{HTML}{009988}
\definecolor{plot6}{HTML}{BBBBBB}

\pgfplotscreateplotcyclelist{colorcycle}{ %
	{plot0, mark=*, thick, mark options=solid},
	{plot1, mark=triangle*, thick, mark options=solid},
	{plot2, mark=square*, thick, mark options=solid},
	{plot3, mark=diamond*, thick, mark options=solid},
	{plot4, mark=pentagon*, thick, mark options=solid},
}
\pgfplotsset{compat=newest}
\usetikzlibrary{positioning,calc,external,spy,shapes}

\DeclareSIUnit{\dBm}{dBm}
\newcommand{\todo}[2][]{\ignorespaces\leavevmode
	\if\relax\detokenize{#1}\relax
	{\color{red}[TODO: #2]}%
	\else
	{\color{red}[TODO (#1): #2]}%
	\fi
}

\theoremstyle{plain}%
\newtheorem{thm}{Theorem}

\newtheorem{lem}{Lemma}
\newtheorem{prop}{Proposition}

\theoremstyle{definition}
\newtheorem{prob}{Problem Statement}
\newtheorem*{prob*}{Problem Statement}

\theoremstyle{remark}
\newtheorem{rem}{Remark}
\newtheorem*{rem*}{Remark}

\newtheoremstyle{example}{\topsep}{\topsep}{}{}{\itshape}{.}{ }{}
\theoremstyle{example}
\newtheorem{example}{Example}
\newtheorem*{example*}{Example}

\addto\extrasenglish{%

}

\algnewcommand\algorithmictry{\textbf{try}}
\algnewcommand\algorithmicendtry{\textbf{end try}}
\algnewcommand\algorithmicexcept{\textbf{except}}
\algblock{Try}{EndTry}
\algcblock[Try]{Try}{Except}{EndIf}
\algrenewtext{Try}{\algorithmictry}
\algrenewtext{EndTry}{\algorithmicendtry}
\algrenewtext{Except}{\algorithmicexcept}

\algnewcommand{\algorithmiclongcomment}[1]{\textcolor{gray}{//~#1}}
\algrenewcommand{\algorithmiccomment}[1]{\hfill\algorithmiclongcomment{#1}}
\algnewcommand{\LongComment}[1]{\algorithmiclongcomment{#1}}

\algrenewcommand{\algorithmicindent}{1em}
\IfPackageLoadedTF{titlesec}{%
	\ifCLASSOPTIONconference%
	\titlespacing{\section}{0pt}{1.5ex plus 1.5ex minus 0.5ex}{0.7ex plus 1ex minus 0ex} %
	\titlespacing{\subsection}{0pt}{1.5ex plus 1.5ex minus 0.5ex}{0.7ex plus .5ex minus 0ex} %
	\else
	\titlespacing{\section}{0pt}{3.0ex plus 1.5ex minus 1.5ex}{0.7ex plus 1ex minus 0ex} %
	\titlespacing{\subsection}{0pt}{3.5ex plus 1.5ex minus 1.5ex}{0.7ex plus .5ex minus 0ex} %
	\fi
	
	\def\thesubsubsectiondis{\arabic{subsubsection})}
	\def\theparagraphdis{\alph{paragraph})}
	\titleformat{\subsubsection}[runin]{\normalfont\normalsize\itshape}{\thesubsubsectiondis}{.5em}{}[:]
	\titlespacing*{\subsubsection}{\parindent}{0ex plus 0.1ex minus 0.1ex}{1ex}
	\titleformat{\paragraph}[runin]{\normalfont\normalsize\itshape}{\theparagraphdis}{.5em}{}[:]
	\titlespacing*{\paragraph}{2\parindent}{0ex plus 0.1ex minus 0.1ex}{1ex}
}{}
\DeclarePairedDelimiter{\abs}{\vert}{\vert}

\DeclarePairedDelimiter{\floor}{\lfloor}{\rfloor}

\newcommand*{\naturals}{\ensuremath{\mathds{N}}}

\newcommand*{\diff}{\ensuremath{\mathrm{d}}}
\newcommand*{\imag}{\ensuremath{\mathrm{j}}}
\newcommand*{\e}{\ensuremath{\mathrm{e}}}

\newcommand*{\unif}{\ensuremath{\mathcal{U}}}

\DeclareMathOperator{\hilbert}{\mathcal{H}}
\newcommand*{\len}{\ensuremath{\ell}}
\newcommand*{\lenref}{\ensuremath{\tilde{\ell}}}
\newcommand*{\htx}[1][]{%
	\if\relax\detokenize{#1}\relax
	\ensuremath{h_{\text{Tx}}}
	\else
	\ensuremath{h_{\text{Tx}, #1}}
	\fi
}
\newcommand*{\hrx}[1][]{%
	\if\relax\detokenize{#1}\relax
	\ensuremath{h_{\text{Rx}}}
	\else
	\ensuremath{h_{\text{Rx}, #1}}
	\fi
}
\newcommand*{\domega}{\ensuremath{\Delta\omega}}
\let\dw\domega
\newcommand*{\dwmax}{\ensuremath{\dw_{\pi}}}
\newcommand*{\dwmin}{\ensuremath{\widetilde{\dw}}}
\newcommand*{\df}{\ensuremath{\Delta f}}
\newcommand*{\dfmax}{\ensuremath{\df_{\pi}}}
\newcommand*{\dfmin}{\ensuremath{\widetilde{\df}}}
\newcommand*{\dmin}{\ensuremath{d_{\text{min}}}}
\newcommand*{\dmax}{\ensuremath{d_{\text{max}}}}
\newcommand*{\dphi}{\ensuremath{\Delta\phi}}
\let\w\omega

\newcommand*{\angleref}{\ensuremath{\alpha}}
\newcommand*{\reflcoeff}{\ensuremath{\Gamma}}
\newcommand*{\constdiel}{\ensuremath{\epsilon}}

\title{An Efficient Frequency Diversity Scheme for Ultra-Reliable Communications in\\Two-Path Fading Channels}
\author{Karl-Ludwig Besser, Eduard A. Jorswieck, and Justin P. Coon
	\thanks{Parts of this work were presented at the 2023 IEEE International Conference on Communications (ICC)~\cite{Besser2023icc} and at the 20th International Symposium on Modeling and Optimization in Mobile, Ad hoc, and Wireless Networks (WiOpt)~\cite{Besser2022wiopt}.}
	\thanks{Karl-Ludwig Besser was with the Institute of Communications Technology, Technische Universit\"at Braunschweig, 38106 Braunschweig, Germany, and is now with the Department of Electrical and Computer Engineering, Princeton University,  Princeton, NJ 08544, USA (email: {karl.besser}@princeton.edu).
	Eduard A. Jorswieck is with the Institute of Communications Technology, Technische Universit\"at Braunschweig, 38106 Braunschweig, Germany (email: {e.jorswieck}@tu-bs.de).
	Justin P. Coon is with the Department of Engineering Science, University of Oxford, Oxford OX1\,3PJ, U.\,K. (email: justin.coon@eng.ox.ac.uk).}
	\thanks{The work of K.-L. Besser is supported by the German Research Foundation (DFG) under grant~BE\,8098/1-1.
	The work of E. Jorswieck is supported in part by the Federal Ministry of Education and Research Germany (BMBF) through the Program of \enquote{Souverän. Digital. Vernetzt.} Joint Project 6G-Research and Innovation Cluster (RIC) under grant~16KISK031.
	The work of J. Coon is supported by the EPSRC under grant number~EP/T02612X/1.}
}

\begin{document}
\maketitle

\begin{abstract}\noindent
	We consider a two-ray ground reflection scenario with unknown distance between transmitter and receiver.
	By utilizing two frequencies in parallel, we can mitigate possible destructive interference and ensure ultra-reliability with only very limited knowledge at the transmitter.
	In order to achieve this ultra-reliability, we optimize the frequency spacing such that the worst-case receive power is maximized.
	Additionally, we provide an algorithm to calculate the optimal frequency spacing.
	Besides the receive power, we also analyze the achievable rate and outage probability.
	It is shown that the frequency diversity scheme achieves a significant improvement in terms of reliability over using a single frequency.
	In particular, we demonstrate the effectiveness of the proposed approach by a numerical simulation of an unmanned aerial vehicle (UAV) flying above flat terrain.
\end{abstract}
\glsresetall

\section{Introduction}\label{sec:introduction}

Reliability is a major requirement for many modern applications of wireless communication systems~\cite{Saad2020,Park2022}.
In particular, this includes autonomous vehicles, e.g., self-driving cars and \glspl{uav}.
It is therefore of great interest to develop techniques, which enable ultra-reliable communications, e.g., assuring low outage probabilities below $10^{-5}$~\cite{Bennis2018}.
This is especially important for scenarios where only limited information, e.g., \gls{csi}, is available at the transmitter, e.g., due to high mobility or in \gls{fdd} systems.

It has been observed that negative dependency between channel gains can significantly improve reliability~\cite{Haber1974,Besser2020twc,Besser2021zoc}.
The basic idea is to establish multi-link diversity and ensure that always one communication link is available, if the others fail.
We will apply this idea in the following to develop a simple frequency diversity scheme that enables ultra-reliable communications in two-ray ground reflection scenarios.
In this two-ray model, it is assumed that only one significant multipath component exists in addition to a \gls{los} connection.
The second component is typically caused by a single reflection on a ground surface.
This could occur in flat outdoor terrain~\cite{Weiler2015}, on large concrete areas, e.g., airports~\cite{Naganawa2017}, and for a \gls{uav} flying above water~\cite{Matolak2014,Matolak2017,Chiu2021}.
It has also been observed that the two-ray model can be appropriate for \gls{v2v} communication scenarios~\cite{Sommer2012,Farzamiyan2020}.
In particular, this includes high frequency bands like \gls{mmwave}~\cite{Guan2017,Khawaja2020,Zochmann2017,Rapaport2018,Jaeckel2017}.

In general, the curvature of the Earth's surface needs to be considered for long-distance outdoor settings and accurate models like the curved-Earth model~\cite{Matolak2015,Matolak2017,Parsons2001} exist.
However, when considering relatively short distances, the flat-Earth model is a valid approximation~\cite{Matolak2017,Parsons2001}, which we adopt throughout this work.

When varying the distance between transmitter and receiver, the relative phase of the two received signal components varies and they may interfere constructively or destructively.
A destructive interference causes a drop of receive power, which in turn could cause an outage of the communication link.
In order to mitigate drops of the signal power on one frequency, a second frequency can be used in parallel.
The use of multiple frequencies in parallel to create diversity and improve the reliability in ground reflection scenarios has already been proposed in~\cite{Haber1974} and~\cite{Berger1972diversity}.
In~\cite{Capriglione2015}, it is analyzed and experimentally verified that frequency diversity improves the performance of distance measurements in outdoor ground reflection scenarios.
Instead of using multiple frequencies in parallel, it was shown experimentally in~\cite{Naganawa2017} that using multiple antennas and carefully choosing the spacing between them can also improve the received power.
A similar problem setup of multipath fading mitigation is considered in~\cite{Basar2021dopplerRIS}, where the authors consider \gls{ris} as solution method.
However, such surfaces need to be deployed first and might not always be available, e.g., in the considered large-scale outdoor scenarios.

{%
The recent trend \gls{jcas} for 6G wireless networks, allows to obtain location information of terminals~\cite{Kwon2021localization}.
This localization information can be exploited as side information to improve the precoding in addition to partial or statistical \gls{csi} at the transmitter.
The usefulness of localization information for transmit optimization depends on the fading channel model.
It holds: the fewer multi-path components, the better.
In particular, at higher frequencies, the multi-path channels become increasingly sparse~\cite{Papasotiriou2018}.
Furthermore, in scenarios without scatterers, but only the ground surface~\cite{Matolak2014dasc}, simple two-path channel models result.
Examples of using the position side information include frequency diversity schemes, opportunistic user scheduling in the time domain~\cite{Viswanath2002}, or spatial selection of antenna positions~\cite{zhu2022modeling}.
}

In this work, we focus on worst-case design for a simple frequency diversity system with very limited information at the communication parties.
We assume that the transmitter does not know the exact distance to the receiver but only has knowledge about lower and upper bounds on the possible distances, e.g., based on a rough estimation of the user positions~{\cite{Kwon2021localization}}. %
In particular, we optimize the spacing between the two frequencies such that the worst-case receive power in the geographical range is maximized.
This can enable ultra-reliability without the need of perfect \gls{csi} at the transmitter.

Our main contributions and the outline of the manuscript are summarized as follows.
\begin{itemize}
	\item We analyze the worst-case received power for a two-ray ground reflection model with unknown distance between transmitter and receiver when employing only a single frequency.\ (\autoref{sec:single-freq})
	\item We analyze the received power and provide a lower bound when employing two frequencies in parallel.\ (\autoref{sec:two-freq})
	\item In particular, we determine the optimal frequency spacing which maximizes the worst-case receive power.\ (\autoref{thm:opt-dw-max-min})
	\item An algorithm and its implementation is provided to calculate the optimal frequency spacing for given system parameters.\ (\autoref{alg:summary-calculation-opt-freq-spacing} and~\cite{BesserGithub})
	\item In addition to the receive power, we also analyze the achievable rate (\autoref{sec:rate-analysis}) and outage probability (\autoref{sec:outage-prob}).
	\item {The benefits of the proposed optimization are highlighted in a numerical example of an \gls{uav} communication system at high 5G~NR frequencies (\autoref{ex:out-prob-uav}).}
\end{itemize}

\section*{Notation}
An overview of the most commonly used variable notation can be found in \autoref{tab:notation}.
\begin{table}[!ht]
\renewcommand*{\arraystretch}{1.2}
\caption{Notation of the Most Commonly Used Variables and System Parameters}%
\label{tab:notation}
\begin{tabularx}{\linewidth}{lX}
	\toprule
	$d$ & Distance between transmitter and receiver (on the ground) [$\si{\meter}$]\\
	$\htx$ & Height of the transmitter [$\si{\meter}$]\\
	$\hrx$ & Height of the receiver [$\si{\meter}$]\\
	$\len$ & Length of the \gls{los} path [$\si{\meter}$]\\
	$\lenref$ & Total length of the reflection path [$\si{\meter}$]\\
	$c$ & Speed of light $=\SI{299792458}{\meter\per\second}$\\
	$\omega = 2\pi f$ & (Angular) frequency ([$\si{\radian\per\second}$]) [$\si{\hertz}$]\\
	$G_{\text{los}}$ & Overall antenna gain for the direction of the \gls{los} path [$\si{\decibel}$]\\
	$G_{\text{ref}}$ & Overall antenna gain for the direction of the reflected path [$\si{\decibel}$]\\
	$\rho=\sqrt{G_{\text{ref}}/G_{\text{los}}}$ & Relative gain for the direction of the reflected path\\
	$\angleref$ & Angle of reflection\\
	$\reflcoeff$ & Reflection coefficient\\
	$\constdiel$ & Dielectric constant of the ground\\
	$P_{t}$ & Transmit power [$\si{\watt}$]\\
	$P_{r}$ & Receive power [$\si{\watt}$]\\
	$\underline{P_{r}}$ & Lower bound of the receive power [$\si{\watt}$]
	\\
	$d_k$ & Distance at which the $k$-th local minimum of the receive power occurs [\si{\meter}]\\
	$\dw=2\pi\df=\omega_2-\omega_1$ & (Angular) frequency spacing ([$\si{\radian\per\second}$]) [$\si{\hertz}$]\\
	$\dwmax$ & First maximum of $\underline{P_r}(d, \dw)$ with respect to $\dw$ [$\si{\radian\per\second}$]\\
	$\dwmin$ & First minimum of $\underline{P_r}(d, \dw)$ with respect to $\dw$ [$\si{\radian\per\second}$]\\
	$R$ & Rate [\si{\bit\per\second}]\\
	$B$ & Bandwidth [\si{\hertz}]\\
	$F$ & Receiver noise figure [\si{\decibel}]\\
	$N_0$ & Noise spectral density [\si{\dBm\per\hertz}]\\
	\bottomrule
\end{tabularx}
\end{table}

In order to simplify the notation, we will omit variables on which functions depend when their value is clear from the context, e.g., we will write $f(x)$ instead of $f(x, y)$ when the value of $y$ is fixed.

Since the angular frequency $\omega=2\pi f$ is a simple scaling of the frequency $f$, we will treat them somewhat interchangeably.
Especially for calculations, it is more convenient to use $\w$, while $f$ is relevant for actual system design.
We will therefore use the frequency $f$ for the numerical examples while expressing all formulas in terms of the angular frequency $\w$.

For a random variable $X$, we use $F_{X}$ and $p_{X}$ for its \gls{cdf} and \gls{pdf}, respectively.
The expectation is denoted by $\mathbb{E}$ and the probability of an event by $\Pr$.
The uniform distribution on the interval $[a,b]$ is denoted as $\unif[a,b]$.
\section{System Model and Problem Formulation}\label{sec:model-problem-formulation}
Throughout the following, we consider the classical two-ray ground reflection model~\cite[Chap.~4.6]{Rappaport2002}.
In this scenario, the propagation environment is approximated as a plane reflecting ground surface.
A single-antenna transmitter is located at height~$\htx$ above the ground.
At distance~$d$, the single-antenna receiver is placed at height~$\hrx$.
This geometrical model is depicted in \autoref{fig:two-ray-model}.

Based on the setup, it can be seen that the transmitted signal is propagated via two separate paths to the receiver.
On the one hand, there exists a \gls{los} propagation with path length~$\len$.
On the other hand, the signal is also reflected by the ground, which leads to the second component.
The total length of the second ray is $\lenref$.
Finally, these two components superimpose at the receiver.

From basic trigonometric considerations, the path lengths can be calculated as
\begin{align}
	\len^2 &= (\htx-\hrx)^2 + d^2\\
	\lenref^2 &= (\htx+\hrx)^2 + d^2\,.
\end{align}

\begin{figure*}
	\begin{equation}\label{eq:rec-power-single-freq}
		P_r(d, \omega; \htx, \hrx, P_t, {\constdiel}) = P_t\left(\frac{c}{2\omega}\right)^2\left(\frac{G_{\text{los}}}{\len^2} + \frac{{{\reflcoeff^2}} G_{\text{ref}}}{\lenref^2} {+} \frac{2{{\reflcoeff}}\sqrt{G_{\text{los}}G_{\text{ref}}}}{\len\lenref}\cos\underbrace{\left[\frac{\omega}{c}\left(\lenref-\len\right)\right]}_{\dphi}\right) %
	\end{equation}
	\hrulefill
\end{figure*}

\begin{figure}
	\centering
	\begin{tikzpicture}
	\node[coordinate] (bottomleft) {};
	\node[right=7 of bottomleft,coordinate] (bottomright) {};
	
	\draw[very thick,plot2,|<->|] (bottomleft) -- node[below]{$d$} (bottomright);
	
	\node[above=4 of bottomleft,label={Transmitter},draw,circle,inner sep=0pt,minimum size=5pt] (anttx) {};
	\draw[thick,|<->|] (bottomleft) -- node[left]{$\htx$} (anttx);
	
	\node[above=2 of bottomright,label={Receiver},draw,circle,inner sep=0pt,minimum size=5pt] (antrx) {};
	\draw[thick,|<->|] (bottomright) -- node[right]{$\hrx$} (antrx);
	
	\node[coordinate] at ($(bottomleft)!0.66!(bottomright)$) (reflection) {};
	
	\draw[plot1,very thick,|<->|] (anttx) -- node[above]{$\len$} (antrx);
	\draw[plot0,very thick,|<->|] (anttx) -- (reflection.center) -- node[above]{$\lenref$} (antrx);

	 \draw[black] ($(reflection)+(1,0)$) arc (0:40.89:1);
	 \draw[black] ($(reflection)+(-1,0)$) arc (180:180-40.89:1);
	 \node[anchor=south east] at ($(reflection)+(.95,.05)$) {$\angleref$};
	 \node[anchor=south west] at ($(reflection)+(-.95,.05)$) {$\angleref$};
\end{tikzpicture}
	\caption{Geometrical model of the considered two-ray ground reflection scenario. The transmitter is placed at height $\htx$ above the ground. The receiver is located at height $\hrx$ at a (ground) distance $d$ away from the transmitter. The \gls{los} path and reflection path have lengths $\len$ and $\lenref$, respectively.
	{The angle of reflection is~$\angleref$.}%
	}
	\label{fig:two-ray-model}
\end{figure}

When transmitting on a single frequency $\w=2\pi f$, the received power~$P_r$ at distance~$d$ is given
in~\eqref{eq:rec-power-single-freq}~\cite[Chap.~4.6]{Rappaport2002}, \cite[Chap.~2.1.2]{Jakes1974}, \cite[Eq.~(2)]{Haber1974} {at the top of this page}
with the additional notation from \autoref{tab:notation}.\footnote{In order to simplify the notation, we will omit variables on which functions depend when their value is clear from the context, e.g., we will write $P_r(d)$ instead of $P_r(d, \omega)$ when the value of $\omega$ is fixed.}

For simplicity, we fix $G_{\text{los}}=1$ throughout the following and use the normalized gain $\rho=\sqrt{G_{\text{ref}}/G_{\text{los}}}$.
This yields
\begin{multline}\label{eq:rec-power-single-freq-params}
	P_r(d, \omega; \htx, \hrx, P_t, \rho, {\constdiel}) = \\
	P_t\left(\frac{c}{2\omega}\right)^2\left(\frac{1}{\len^2} + \frac{{{\reflcoeff^2}}\rho^2}{\lenref^2} {{+}} \frac{2{{\reflcoeff}}\rho}{\len\lenref}\cos\underbrace{\left[\frac{\omega}{c}\left(\lenref-\len\right)\right]}_{\dphi}\right)\,.
\end{multline}
Additionally, we assume that the gain on the reflected path is less than on the direct path, i.e., $\rho\leq 1$, e.g., due to additional absorption on the ground.
{%
The reflection coefficient~$\reflcoeff$ is given as~\cite[{Eq.~(2.1-3)}]{Jakes1974}
\begin{equation}\label{eq:def-refl-coeff}
	\reflcoeff(\angleref) = \frac{\sin\angleref - z}{\sin\angleref + z}\,,
\end{equation}
with
\begin{equation}
	z = \sqrt{\constdiel-\cos^2\angleref}\,,
\end{equation}
where $\constdiel>1$ is the dielectric constant of the ground, and $\angleref$ is the angle of reflection, cf.~\autoref{fig:two-ray-model}.
Since the reflection coefficient depends on the angle of reflection, it depends on the distance~$d$ between receiver and transmitter.
This behavior is shown for different values of the dielectric constant of the ground~$\constdiel$ in \autoref{fig:refl-coeff}.
Typical values range from $\constdiel=4$ (poor ground) to $\constdiel=81$ (water)~\cite[{Chap.~2, Tab.~1}]{Jakes1974}.
It can be seen that the reflection coefficient approaches $-1$ for large distances~$d$ for all $\constdiel$.
Similarly, it approaches the finite negative value $\frac{1-\sqrt{\constdiel}}{1+\sqrt{\constdiel}}$ for small~$d$.
It should be noted that the ideal reflecting ground would have $\rho=1$ and $\reflcoeff=-1$ independent of the angle of reflection, i.e., independent of the distance~$d$.
\begin{figure}
	\centering
	\begin{tikzpicture}
\begin{axis}[
	width=.96\linewidth,
	height=.23\textheight,
	xlabel={Distance~$d$ [\si{\meter}]},
	ylabel={Reflection Coefficient~$\reflcoeff$},
	xlabel near ticks,
	ylabel near ticks,
	xmin=1,
	xmax=1000,
	xmode=log,
	ymin=-1,
	ymax=0,
	cycle list name=colorcycle,
	ymajorgrids,
	xmajorgrids,
	xminorgrids,
	grid style={line width=.1pt, draw=gray!20},
	major grid style={line width=.25pt,draw=gray!30},
	legend pos=north east,
	legend cell align=left,
	]
	\addplot+[mark repeat=50] table {data/refl_coeff-2.400000E+09-t10.0-r1.5-eps2.00.dat};
	\addlegendentry{$\constdiel=2$};
	
	\addplot+[mark repeat=50] table {data/refl_coeff-2.400000E+09-t10.0-r1.5-eps20.00.dat};
	\addlegendentry{$\constdiel=20$};
	
	\addplot+[mark repeat=50] table {data/refl_coeff-2.400000E+09-t10.0-r1.5-eps80.00.dat};
	\addlegendentry{$\constdiel=80$};
\end{axis}
\end{tikzpicture}
	\caption{Reflection coefficient~$\reflcoeff$ from \eqref{eq:def-refl-coeff} for different dielectric constants of the ground~$\constdiel$.}%
	\label{fig:refl-coeff}
\end{figure}
}

It is well-known that the two components can interfere constructively or destructively at the receiver, depending on the distance~$d$.
This leads to local minima in the receive power at certain distances, which in turn can lead to outages in the transmission.
In order to mitigate these drops in the receive power, we propose to use a second frequency in parallel.
The exact problem formulation is described in the following.

\subsection{Problem Formulation}
Throughout the following, we will consider a two-ray ground reflection scenario where the height of the transmitter $\htx$, the height of the receiver $\hrx$, and the base frequency $\omega_1$ are fixed.
In contrast, the distance between transmitter and receiver $d$ varies and is unknown.
Only the range of possible distances $d$ is known, i.e., $d\in[\dmin, \dmax]$.

In order to ensure a high reception quality at any distance within the interval, the transmitter employs a second frequency~$\omega_2 = \omega_1+\domega$ such that we can compensate for possible destructive interference of the two rays.
This leads to the following problem.

\begin{prob}\label{prob:opt-prob}
	Since the transmitter only knows the range of $d$, i.e., that $d\in[\dmin, \dmax]$, we adjust the frequency spacing~$\domega$ such that the \emph{worst-case} receive power in $[\dmin, \dmax]$ is maximized.
	This optimization problem can be formulated as
	\begin{equation}\label{eq:opt-problem}
		\max_{\domega}\min_{d\in[\dmin, \dmax]} P_{r,1}(d, \omega_1) + P_{r,2}(d, \omega_2)\,,
	\end{equation}
	where $P_{r,1}(d, \omega_1)$ is the received power on the first frequency~$f_1=\omega_1/(2\pi)$ at distance~$d$ and $P_{r,2}(d, \omega_2)$ is the received power for the second frequency~$f_2 = \omega_2/(2\pi) = (\omega_1+\dw)/(2\pi)$ at distance~$d$.
\end{prob}
\section{Single Frequency}\label{sec:single-freq}
In order to solve \eqref{eq:opt-problem}, we first need to analyze the receive power for a single frequency~$\w$.
Since we are interested in a worst-case design, we analyze the worst-case receive power of a single frequency in the following.
In order to do this, we start by investigating the distances at which destructive interference occurs.

\subsection{Destructive Interference}
From~\eqref{eq:rec-power-single-freq}, it can be seen that we get a (local) minimum of the receive power when the direct and reflected signals interfere destructively~\cite{Loyka2001}.
This occurs whenever the phase difference~$\dphi$ is a multiple of $2\pi$, i.e.,
\begin{equation}\label{eq:condition-phase-shift-minimum-single-freq}
	\dphi = \frac{\omega}{c}\left(\lenref-\len\right) = 2\pi k, \quad k\in\naturals_0\,.
\end{equation}
It can easily be verified that $\dphi$ is a decreasing function in $d$ and it therefore follows that
\begin{align*}
	\dphi_{\text{max}}
	&= \lim\limits_{d\to 0} \dphi %
	= \frac{2\omega\min\{\htx,\hrx\}}{c}
	\;\;\text{and}\;\; \lim\limits_{d\to\infty} \dphi = 0\,.
\end{align*}
This shows that $\dphi$ decreases from a finite value~$\dphi_{\text{max}}$ to $0$.
Hence, there always exists a finite number of multiples of $2\pi$ with $k_{\text{max}} = \floor{\frac{\dphi_{\text{max}}}{2\pi}}$, i.e., there exist $k_{\text{max}}$~local minima of the receive power.

The distance~$d_k$ at which the $k$-th minimum occurs, is given by solving~\eqref{eq:condition-phase-shift-minimum-single-freq} as
\begin{equation}\label{eq:distance-k-min-single-freq}
	d_k^2(\omega) = \frac{\left({(c \pi k)^2 - (\omega \hrx)^2}\right) \left({(c \pi k)^2 - (\omega \htx)^2}\right)}{(\omega c \pi k)^2}\,,%
\end{equation}
with $k=1, \dots{}, k_{\text{max}}$.

\begin{example}\label{ex:single-freq-local-min}
	An illustration of $\dphi$ can be found in \autoref{fig:phase-shift-single-freq}.
	The parameters are set to $\w/c=10$ ($f=\SI{477}{\mega\hertz}$), $\htx=\SI{10}{\meter}$, and $\hrx=\SI{1.5}{\meter}$.
	Additionally, we indicate the distances~$d_k$.
	Since $k_{\text{max}}=4$, there exist four~$d_k$ at which a local minimum occurs.
	For the selected parameters, they are evaluated to $d_1=\SI{46.7}{\meter}$, $d_2=\SI{21.6}{\meter}$, $d_3=\SI{12.3}{\meter}$, and $d_4=\SI{6.5}{\meter}$.
	The corresponding received power from~\eqref{eq:rec-power-single-freq} is shown in \autoref{fig:rec-power-single-freq}.
	It can be seen that a minimum occurs at $d=d_k$, with the lowest minimum being at the smallest~$k$, i.e., $k=1$, which corresponds to the highest distance of all $d_k$.
	For comparison, we additionally show the received power for $f_2=\SI{2.4}{\giga\hertz}$ in \autoref{fig:rec-power-single-freq}.
	\begin{figure}
		\centering
		\begin{tikzpicture}%
	\begin{axis}[
		width=.97\linewidth,
		height=.28\textheight,
		xlabel={Distance $d$ [\si{\meter}]},
		ylabel={Phase Shift $\Delta\phi$},
		xlabel near ticks,
		ylabel near ticks,
		xmin=1,
		xmax=1000,
		xmode=log,
		ymin=0,
		cycle list name=colorcycle,
		extra x ticks={46.66, 21.64, 12.33, 6.47},
		extra x tick labels={$d_1$, $d_2$, {}, $d_4$},
		extra y ticks={6.28, 12.57, 18.85, 25.13},
		extra y tick labels={$2\pi$, $4\pi$, $6\pi$, $8\pi$},
		extra x tick style={grid=none,font=\small},
		extra y tick style={grid=none,font=\small},
		ymajorgrids,
		xmajorgrids,
		xminorgrids,
		grid style={line width=.1pt, draw=gray!20},
		major grid style={line width=.25pt,draw=gray!30},
		]

		\addplot+[domain=1:1000] {10*(sqrt(132.25+x^2)-sqrt(72.25+x^2))};
		\addplot[black,thick,dashed] coordinates {(1, 6.28) (46.66, 6.28) (46.66, 0)};
		\addplot[black,thick,dashed] coordinates {(1, 12.57) (21.64, 12.57) (21.64, 0)};

		\addplot[black,thick,dashed] coordinates {(1, 18.85) (12.33, 18.85) (12.33, 0)};
		\addplot[black,thick,dashed] coordinates {(1, 25.13) (6.47, 25.13) (6.47, 0)};
	\end{axis}
\end{tikzpicture}
		\caption{%
			Relative phase shift $\dphi$ from \eqref{eq:condition-phase-shift-minimum-single-freq} for $\omega/c=10$, $\htx=\SI{10}{\meter}$, and $\hrx=\SI{1.5}{\meter}$.
			Additionally the distances $d_k$, $k=1, \dots{}, 4$, from \eqref{eq:distance-k-min-single-freq} are indicated.
			(\autoref{ex:single-freq-local-min})
		}
		\label{fig:phase-shift-single-freq}
	\end{figure}
\end{example}

\subsection{Worst-Case Receive Power}
When $d$ is anywhere in the interval $[\dmin, \dmax]$, the lowest drop is at the smallest~$k$ such that $d_k$ is still in $[\dmin, \dmax]$.
However, in order to determine the global minimum of the receive power in $[\dmin, \dmax]$, the boundary points~$\dmin$ and $\dmax$ need to be taken into account.
This leads to the following result of the minimal receive power when only a single frequency is used.

\begin{thm}[Minimal Receive Power (Single Frequency)]\label{thm:min-rec-power-single-freq}
	Consider the described two-ray ground reflection model with a single frequency~$\omega=2\pi f$.
	The distance~$d$ between transmitter and receiver is in the interval $[\dmin, \dmax]$.
	The minimal receive power is then given as
	\begin{multline}
		\label{eq:min-rec-power-single-freq}
		\min_{d\in[\dmin, \dmax]} P_{r}(d) = \\
		\min \left\{P_r(\dmin), P_r(\dmax), P_r\left(\max_{d_k\in[\dmin, \dmax]} d_k\right)\right\}\,.
	\end{multline}
\end{thm}

\begin{example}[Single Frequency Worst-Case Receive Power]\label{ex:single-freq-worst-case}
	For a numerical example, we take the parameters that are used in \autoref{ex:single-freq-local-min} and additionally compare it to a higher frequency scenario.
	In particular, we fix $\htx=\SI{10}{\meter}$, $\hrx=\SI{1.5}{\meter}$, and $P_t=1$.
	For this example, we assume a perfect reflection on the ground without any additional absorption, i.e., $\rho=1$ and $\reflcoeff=-1$.
	The receiver is assumed to be randomly located at a distance between $\dmin=\SI{30}{\meter}$ and $\dmax=\SI{100}{\meter}$ from the transmitter.
	
	For the lower frequency~$\omega_1=10c$, i.e., $f_1=\SI{477}{\mega\hertz}$, we get $P_r(\dmin, \omega_1)=\SI{-50}{\decibel}$, $P_r(\dmax, \omega_1)=\SI{-60}{\decibel}$, and $P_r(d_1(\omega_1), \omega_1)=\SI{-97}{\decibel}$ with $d_1(\omega_1)=\SI{46.7}{\meter}$.
	Based on~\eqref{eq:min-rec-power-single-freq} from \autoref{thm:min-rec-power-single-freq}, we determine that the worst-case receive power is equal to \SI{-97}{\decibel}.
	
	In contrast, for a higher frequency~$f_2=\SI{2.4}{\giga\hertz}$, there are multiple local minima at locations~$d_k$, which lie in the interval~$[\dmin, \dmax]$.
	According to~\eqref{eq:min-rec-power-single-freq}, we need to determine the maximum of all~$d_k\in[\dmin, \dmax]$.
	For the considered parameters, this is calculated to $d_3(\omega_2)=\SI{79.4}{\meter}$ with $P_r(d_3(\omega_2), \omega_2)=\SI{-125}{\decibel}$.
	The received powers at the boundary points are evaluated to $P_r(\dmin, \omega_2)=\SI{-64}{\decibel}$ and $P_r(\dmax, \omega_2)=\SI{-75}{\decibel}$.
	Hence, the worst-case receive power when using only frequency~$f_2$ is $\SI{-125}{\decibel}$.
	\begin{figure}
		\centering
		\begin{tikzpicture}%
	\begin{axis}[
		width=.93\linewidth,
		height=.27\textheight,
		xlabel={Distance $d$ [\si{\meter}]},
		ylabel={Received Power $P_r$ [\si{\decibel}]},
		xlabel near ticks,
		ylabel near ticks,
		xmin=1,
		xmax=1000,
		xmode=log,
		ymax=-40, %
		ymin=-130, %
		legend pos=south west,
		legend cell align=left,
		cycle list name=colorcycle,
		extra x ticks={46.66, 21.64},
		extra x tick labels={$d_1$, $d_2$},
		extra x tick style={grid=none},
		extra y tick style={grid=none},
		ymajorgrids,
		xmajorgrids,
		xminorgrids,
		grid style={line width=.1pt, draw=gray!20},
		major grid style={line width=.25pt,draw=gray!30},
		]
		\addplot+[mark repeat=60] table[x=distance, y=power] {data/power_single-4.771345E+08-t10.0-r1.5.dat};
		\addlegendentry{$f_1=\SI{477}{\mega\hertz}$};
		
		\addplot+[mark repeat=80] table[x=distance, y=power] {data/power_single-2.400000E+09-t10.0-r1.5.dat};
		\addlegendentry{$f_2=\SI{2.4}{\giga\hertz}$};
		
		\addplot[black,dashed,thick] coordinates {(46.66, -40) (46.66, -130)};
		\addplot[black,dashed,thick] coordinates {(21.64, -40) (21.64, -130)};
	\end{axis}
\end{tikzpicture}
		\caption{Received power~$P_{r}(d)$ from~\eqref{eq:rec-power-single-freq} when using a single frequency~$f$ with system parameters $\htx=\SI{10}{\meter}$, $\hrx=\SI{1.5}{\meter}$, $\rho=1$, {$\reflcoeff=-1$}, and $P_t=1$ for $f=f_1=\SI{477}{\mega\hertz}$ and $f=f_2=\SI{2.4}{\giga\hertz}$. Additionally, the distances~$d_1(\omega_1)=\SI{46.7}{\meter}$ and $d_2(\omega_1)=\SI{21.6}{\meter}$ from \eqref{eq:distance-k-min-single-freq} are indicated. (\autoref{ex:single-freq-local-min} and \autoref{ex:single-freq-worst-case})}
		\label{fig:rec-power-single-freq}
	\end{figure}
\end{example}

\begin{example}[{Single Frequency Receive Power with Different Gains {and Reflection Coefficients}}]\label{ex:single-freq-rho}
	In the previous example, we assumed that the antenna and reflection gain on the reflected path is the same as the one on the \gls{los} path, i.e., $\rho=1$.
	In practical systems, it is likely that $\rho < 1$, e.g., due to an additional absorption at the ground.
	Therefore, we now compare the receive power for different values of $\rho$ in \autoref{fig:rec-power-single-freq-rho}.
	The system parameters are again set to $f=\SI{2.4}{\giga\hertz}$, $\htx=\SI{10}{\meter}$, $\hrx=\SI{1.5}{\meter}$, $P_t=1$, $\dmin=\SI{30}{\meter}$, and $\dmax=\SI{100}{\meter}$.
	In order to isolate the influence of the parameter~$\rho$, we still fix the reflection coefficient to $\reflcoeff=-1$.

	For small~$\rho$, the influence of the reflected ray reduces, and the total receive power is approximately determined by the path-loss of the \gls{los} component.
	In this case, the destructive interference only has a negligible effect.
	For $\rho=0.1$, the receive power at $d_3$ is around $\SI{-79}{\decibel}$.
	At the boundaries of the distance interval, we get $P_r(\dmin)=\SI{-69.2}{\decibel}$ and $P_r(\dmax)=\SI{-79.4}{\decibel}$.
	Thus, the worst-case receive power in $[\dmin, \dmax]$ is given by $P_r(\dmax)$.
	
	In contrast, for a larger value of $\rho=0.5$, the destructive interference is significant and we evaluate the receive power at $d_3$ to $\SI{-84.1}{\decibel}$.
	This is also the worst-case receive power in the interval $[\dmin, \dmax]$ for $\rho=0.5$.
	Similarly, the worst-case receive power for $\rho=1$ is also given by the local minimum at $d_3$ with $P_r(d_3)=\SI{-125}{\decibel}$, cf. \autoref{ex:single-freq-worst-case}.
	It should also be noted that the constructive interference is more significant for large values of $\rho$.

	Next, we investigate the influence of a varying reflection coefficient from~\eqref{eq:def-refl-coeff}.
	For this, we fix $\rho=1$ and show the receive power~$P_r$ for different dielectric constants~$\constdiel$ in \autoref{fig:rec-power-single-freq-gamma}.
	Similarly to the influence of $\rho$, it can be seen that the destructive interference gets less severe with decreasing~$\constdiel$.
	As the absolute value of the reflection coefficient~$\reflcoeff$ is smaller for small distances, cf.~\autoref{fig:refl-coeff}, the receive power shows less fluctuation for small distances.
	This effect is more noticeable for small~$\constdiel$.
	The severity of the drops in receive power due to destructive interference is determined by the absolute value of the product of $\rho$ and $\reflcoeff$, since values close to zero indicate a weak signal from the reflected path.
	The worst case in terms of destructive interference therefore occurs for larger values of~$\abs{\rho\reflcoeff}$, which is discussed in more detail in the following proposition.
	
	\begin{figure}
		\centering
		\subfigure[{Influence of the relative gain~$\rho$ ({$\reflcoeff=-1$})\label{fig:rec-power-single-freq-rho}}]{\centering\begin{tikzpicture}%
	\begin{axis}[
		width=.93\linewidth,
		height=.27\textheight,
		xlabel={Distance $d$ [\si{\meter}]},
		ylabel={Received Power $P_r$ [\si{\decibel}]},
		xlabel near ticks,
		ylabel near ticks,
		xmin=10,
		xmax=1000,
		xmode=log,
		ymax=-40, %
		ymin=-130, %
		legend pos=north east,
		legend cell align=left,
		cycle list name=colorcycle,
		extra x tick style={grid=none},
		extra y tick style={grid=none},
		ymajorgrids,
		xmajorgrids,
		xminorgrids,
		grid style={line width=.1pt, draw=gray!20},
		major grid style={line width=.25pt,draw=gray!30},
		]
		\addplot+[mark repeat=80] table[x=distance, y=power] {data/power_single-2.400000E+09-t10.0-r1.5-rho1.00.dat};
		\addlegendentry{$\rho=1$};
		\addplot+[mark repeat=100] table[x=distance, y=power] {data/power_single-2.400000E+09-t10.0-r1.5-rho0.50.dat};
		\addlegendentry{$\rho=0.5$};
		\addplot+[mark repeat=130] table[x=distance, y=power] {data/power_single-2.400000E+09-t10.0-r1.5-rho0.10.dat};
		\addlegendentry{$\rho=0.1$};
	\end{axis}
\end{tikzpicture}}
		
		{\subfigure[{Influence of the reflection coefficient and dielectric constant~$\constdiel$ ($\rho=1$)\label{fig:rec-power-single-freq-gamma}}]{\begin{tikzpicture}
\begin{axis}[
	width=.93\linewidth,
	height=.27\textheight,
	xlabel={Distance $d$ [\si{\meter}]},
	ylabel={Received Power $P_r$ [\si{\decibel}]},
	xlabel near ticks,
	ylabel near ticks,
	xmin=10,
	xmax=1000,
	xmode=log,
	ymax=-40, %
	ymin=-130, %
	legend pos=north east,
	legend cell align=left,
	cycle list name=colorcycle,
	extra x tick style={grid=none},
	extra y tick style={grid=none},
	ymajorgrids,
	xmajorgrids,
	xminorgrids,
	grid style={line width=.1pt, draw=gray!20},
	major grid style={line width=.25pt,draw=gray!30},
	]

	\addplot+[mark repeat=130,very thick] table[x=distance, y=power] {data/power_single-2.400000E+09-t10.0-r1.5-rho1.00-eps80.00.dat};
	\addlegendentry{$\constdiel=80$};
	\addplot+[mark repeat=100] table[x=distance, y=power] {data/power_single-2.400000E+09-t10.0-r1.5-rho1.00-eps20.00.dat};
	\addlegendentry{$\constdiel=20$};
	\addplot+[mark repeat=150] table[x=distance, y=power] {data/power_single-2.400000E+09-t10.0-r1.5-rho1.00-eps2.00.dat};
	\addlegendentry{$\constdiel=2$};
\end{axis}
\end{tikzpicture}}}
		\caption{Received power $P_{r}(d)$ from \eqref{eq:rec-power-single-freq} when using a single frequency $f=\SI{2.4}{\giga\hertz}$ with system parameters $\htx=\SI{10}{\meter}$, $\hrx=\SI{1.5}{\meter}$, and $P_t=1$. (\autoref{ex:single-freq-rho})}
		\label{fig:rec-power-single-parameter-influence}
	\end{figure}
\end{example}

\begin{prop}[{Worst-Case $\rho$ {and $\reflcoeff$}}]\label{prop:worst-case-rho}
	The destructive interference is getting worse for decreasing~$\rho\reflcoeff < 0$ and attains a minimum at $\rho\reflcoeff = -\lenref/\len < -1$.
	However, since we have $-1 < \reflcoeff < 0$ and $0\leq\rho\leq 1$, the worst-case destructive interference of the two rays is given for $\rho{\reflcoeff} = -1$.
	
	It should be noted that this is independent of the distance, i.e., we replace the reflection coefficient~$\reflcoeff(\angleref)$ by its worst-case $\reflcoeff=-1$ independent of the angle of reflection.
\end{prop}
\begin{proof}
	The proof can be found in \autoref{app:proof-prop-worst-case-params}.
\end{proof}

Based on the discussion in \autoref{prop:worst-case-rho}, we fix {$\rho\reflcoeff=-1$} throughout the rest of this work.
\section{Two Frequencies}\label{sec:two-freq}
Since the drops in receive power due to destructive interference of the two rays cannot be avoided without knowledge of~$d$ when a single frequency is used, we will now employ a second frequency to mitigate these minima.
As described in \autoref{prob:opt-prob}, we aim to optimize the frequency spacing~$\df$ for a second frequency~$f_2=f_1+\df=\omega_2/(2\pi)=(\omega_1+\dw)/(2\pi)$ given a base frequency~$f_1=\omega_1/(2\pi)$.

As discussed in \autoref{prop:worst-case-rho}, the destructive interference is worst for $\rho{\reflcoeff}=-1$.
Therefore, we will only consider this case throughout this section.

As an additional assumption for this work, we consider ideal \gls{mrc} of the signals on the two frequencies.
While issues of combining signals with different symbols, pulses, and potentially modulation formats are relevant and important for the implementation of such systems, they are beyond of the scope of this work.
This also includes synchronization issues, which could be addressed through the use of \gls{ofdm}.
When \gls{ofdm} is employed, the receiver will perform a rapid synchronization, typically utilizing the Schmidl-Cox method~\cite{Schmidl1997}.
With this, the start of the frame and the beginning of the symbol can be found, and carrier frequency offsets of many subchannel spacings can be corrected.
This implies that we can coherently combine the symbols of different carriers by \gls{mrc}, because we know the start and stop of the \gls{ofdm} symbol and have the correct carrier frequency information.

With this assumption, the received power at distance~$d$ is given as the sum power~$P_{r,1}(d, \omega_1)+P_{r,2}(d, \omega_2)$.
For a fair comparison with the single frequency case, we assume that the total transmit power~$P_t$ remains the same.
Thus, for the first frequency, $\theta P_t$ with $\theta\in[0, 1]$, is used as the transmit power and $\bar{\theta} P_t$, $\bar{\theta}=1-\theta$, for the second.
This leads to the expression of the total received power in~\eqref{eq:rec-power-two-freq} at the bottom of this page.
\begin{figure*}[b]
	\hrulefill
\begin{equation}\label{eq:rec-power-two-freq}%
	P_{r} = {P_t}\left(\frac{c}{2}\right)^2 \left[\left(\frac{\theta}{\omega_1^2}+\frac{\bar{\theta}}{\omega_2^2}\right)\left(\frac{1}{\len^2}+\frac{1}{\lenref^2}\right)-\frac{2}{\len\lenref}\left(\frac{\theta\cos\left(\frac{\omega_1}{c}(\lenref-\len)\right)}{\omega_1^2} + \frac{\bar{\theta}\cos\left(\frac{\omega_2}{c}(\lenref-\len)\right)}{\omega_2^2}\right)\right] %
\end{equation}
\begin{equation}
\label{eq:rec-power-sum-lower-bound}
	\underline{P_{r}}(d, \dw; \omega_1, \htx, \hrx, P_t, \theta) =
	{P_t} \left(\frac{c}{2}\right)^2 \left[ 
	\left(\frac{\theta}{\omega_1^2}+\frac{\bar{\theta}}{\omega_2^2}\right) \left(\frac{1}{\len^2}+\frac{1}{\lenref^2}\right) \\ -\frac{2}{\len\lenref}\sqrt{\left(\frac{\theta}{\omega_1^2}\right)^2 + \left(\frac{\bar{\theta}}{\omega_2^2}\right)^2 + \frac{2\theta\bar{\theta}\cos\left(\frac{\domega}{c}(\lenref-\len)\right)}{\omega_1^2 \omega_2^2}}
	\right]%
\end{equation}
\end{figure*}

Since we are particularly interested in improving the worst-case performance, we will consider a lower bound on~$P_r$ in the following.
\begin{lem}[Lower Bound on the Sum Power for Two Frequencies]\label{lem:rec-power-lower-bound-two-freq}
	For the described two-ray ground reflection system using two frequencies~$\omega_1, \omega_2$ in parallel, the received (sum) power~$P_r$, is lower bounded by 
	\eqref{eq:rec-power-sum-lower-bound} at the bottom of this page.
\end{lem}
\begin{proof}
	The proof can be found in \autoref{app:proof-rec-power-lower-bound}.
\end{proof}

\begin{prop}[Optimal Power Split for Two Frequencies]\label{prop:opt-power-split}
	In general, the optimal power split~$\theta^\star$, which maximizes~$P_r$, depends on all of the parameters~$\len$, $\lenref$, $\w_1$, and $\dw$.
	Hence, the optimal power split varies with varying distance~$d$.
	However, for base frequencies~$\w_1$ which are large compared to the frequency spacing~$\dw$, $\theta^\star$ approaches~\num{0.5}, independent of $d$.
\end{prop}

Since we do not assume exact knowledge about~$d$ at the transmitter, we cannot adjust the power split~$\theta$ to be the exact optimum~$\theta^\star$.
However, in most communication systems, we typically have the case that $\w_1\gg\dw$.
Therefore, the approximation of equal power allocation, i.e., $\theta=0.5$, from \autoref{prop:opt-power-split} is a reasonable strategy, which we will use throughout the following.

\begin{example}[Sum Power Lower Bound]\label{ex:sum-power-lower-bound}
As an example, we show the received (sum) power from~\eqref{eq:rec-power-two-freq} and the lower bound from~\eqref{eq:rec-power-sum-lower-bound} for $f_1=\SI{2.4}{\giga\hertz}$, $\df=\SI{250}{\mega\hertz}$, $\htx=\SI{10}{\meter}$, and $\hrx=\SI{1.5}{\meter}$ in \autoref{fig:rec-power-two-freq}.
It can clearly be seen that the lower bound is the lower envelope of the actual received power.
Similar to the case that only a single frequency is used, the received power varies with the distance~$d$ and shows both minima and maxima.
While the actual received power~$P_r$ oscillates at a high frequency over the distance~$d$, the (spatial) frequency of the lower bound~$\underline{P_r}$ is determined by the difference in frequencies~$\df=\dw/(2\pi)$.
\begin{figure}
	\centering
	\begin{tikzpicture}%
	\begin{axis}[
		width=.93\linewidth,
		height=.27\textheight,
		xlabel={Distance $d$ [\si{\meter}]},
		ylabel={Received Power $P_r$ [\si{\decibel}]},
		xlabel near ticks,
		ylabel near ticks,
		xmin=1,
		xmax=1000,
		xmode=log,
		ymax=-50,
		ymin=-120,
		cycle list name=colorcycle,
		legend cell align=left,
		legend pos=south west,
		extra x tick style={grid=none},
		extra y tick style={grid=none},
		ymajorgrids,
		xmajorgrids,
		xminorgrids,
		grid style={line width=.1pt, draw=gray!20},
		major grid style={line width=.25pt,draw=gray!30},
		]
		\addplot+[mark repeat=40] table[x=distance, y=powerSum] {data/power_sum-2.400000E+09-df2.500000E+08-t10.0-r1.5.dat};
		\addlegendentry{Sum Power $P_r$};
		\addplot+[mark repeat=40, dashed] table[x=distance, y=envelope] {data/power_sum-2.400000E+09-df2.500000E+08-t10.0-r1.5.dat};
		\addlegendentry{Lower Bound $\underline{P_r}$};

	\end{axis}
\end{tikzpicture}
	\caption{Received power for two parallel frequencies with $f_1=\SI{2.4}{\giga\hertz}$, $\df=\SI{250}{\mega\hertz}$, $\htx=\SI{10}{\meter}$, and $\hrx=\SI{1.5}{\meter}$. Both the actual value~$P_r$ from~\eqref{eq:rec-power-two-freq} and the lower bound~$\underline{P_r}$ from~\eqref{eq:rec-power-sum-lower-bound} are shown. (\autoref{ex:sum-power-lower-bound})}%
	\label{fig:rec-power-two-freq}
\end{figure}
\end{example}

In the following, we give an approximation for the locations of the local maxima and minima of $\underline{P_r}$ when two frequencies are used in parallel.

\begin{lem}\label{lem:approx-dw-max-min-sum-power-envelope}
	For $\omega_1\gg\dw$, the bound on the received power~$\underline{P_r}$ from~\eqref{eq:rec-power-sum-lower-bound} has a (local) maximum approximately at
	\begin{equation}\label{eq:dw-maximum}
		\dw_{\pi,k} = \frac{\pi c}{\lenref-\len} (2k+1), \quad k\in{\naturals_0}\,,
	\end{equation}
	and a (local) minimum approximately at
	\begin{equation}\label{eq:dw-minimum}
		\widetilde{\domega_k} = \frac{2\pi c}{\lenref-\len}k, \quad k\in\naturals_0\,.
	\end{equation}
\end{lem}
\begin{proof}
	For $\omega_1\gg\dw$, we can use the approximation that $\omega_1\approx\omega_2=\omega_1+\dw$.
	With this, $\underline{P_r}$ from~\eqref{eq:rec-power-sum-lower-bound} can be simplified to
	\begin{equation*}
		\underline{P_r}\approx \frac{P_t}{\omega_1^2}\left(\frac{c}{2}\right)^2 \left[\left(\frac{1}{\len^2}+\frac{1}{\lenref^2}\right)-\frac{2}{\len\lenref}\abs*{\cos\frac{\dphi}{2}}\right]
	\end{equation*}
	with $\dphi = \frac{\dw}{c}(\lenref-\len)$.
	From this, it can directly be seen that $\underline{P_r}$ has a local maximum at $\dphi/2 = (2k+1)\pi/2$ and a local minimum at $\dphi/2 = k\pi$ with $k\in\naturals_0$.
	Solving this for $\dw$, we obtain $\dwmax$ and $\dwmin$ from \eqref{eq:dw-maximum} and \eqref{eq:dw-minimum}, respectively.
\end{proof}

\begin{rem}
	The important consequence of \autoref{lem:approx-dw-max-min-sum-power-envelope} is that for large~$\omega_1$, $\underline{P_r}(d, \dw)$ is an increasing function in~$\dw$ for $\widetilde{\dw_{k-1}}<\dw<\dw_{\pi,k}$ and decreasing for $\dw_{\pi,k} < \dw < \widetilde{\dw_{k}}$, with $k=1, 2, \dots{}$.
	Also note that we always have ${\widetilde{\domega_0}=0}$, i.e., there is a local minimum at~$\dw=0$, which corresponds to the single frequency case.
\end{rem}

Throughout the following, we will refer to $\dw_{\pi,{1}}$ and $\widetilde{\domega_1}$ from~\eqref{eq:dw-maximum} and~\eqref{eq:dw-minimum} simply as $\dwmax$ and $\dwmin$, respectively.

\begin{example}[Approximation of the Peak Positions]\label{ex:approx-peak-positions}
	A numerical illustration of the approximations from \autoref{lem:approx-dw-max-min-sum-power-envelope} can be found in \autoref{fig:rec-power-df-approximation-peaks}.
	For the example, we consider two different base frequencies~$f_1$, namely $f_1^{(1)}=\SI{100}{\mega\hertz}$ and $f_1^{(2)}=\SI{2.4}{\giga\hertz}$.
	The remaining parameters are fixed to $d=\SI{50}{\meter}$, $\htx=\SI{10}{\meter}$, and $\hrx=\SI{1.5}{\meter}$.
	Besides the lower bound on the power~$\underline{P_r}$ from \eqref{eq:rec-power-sum-lower-bound}, we show the locations of the first peak~$\dwmax$ and drop~$\dwmin$.
	The approximated values are given by~\eqref{eq:dw-maximum} and~\eqref{eq:dw-minimum} in \autoref{lem:approx-dw-max-min-sum-power-envelope}.
	For comparison, we additionally indicate the values of the exact minimum and maximum locations, which are determined numerically~\cite{BesserGithub}.
	First, it can clearly be seen that the approximation of the minimum location~$\dwmin$ is very close to the exact value for both base frequencies~$f_1$.
	Both the approximation and the exact values are evaluated to around $\dfmin=\dwmin/(2\pi)=\SI{510}{\mega\hertz}$.
	
	In contrast, the approximation of the location of the first maximum~$\dwmax$ becomes more accurate for large~$f_1$.
	For the considered parameter values, the approximation from~\eqref{eq:dw-maximum} is evaluated to $\dfmax=\dwmax/(2\pi)=\SI{255}{\mega\hertz}$.
	For $f_1^{(1)}$, the maximum actually occurs at around $\df=\SI{178}{\mega\hertz}$, which corresponds to a relative error of around~\SI{43}{\percent}.
	However, for the larger base frequency~$f_1^{(2)}$, the exact location of the maximum is at around \SI{253}{\mega\hertz}.
	In this case, the relative error of the approximation is only around~\SI{0.79}{\percent}.
	This indicates that the approximation is accurate enough for practical purposes when considering typical base frequencies of above~\SI{2}{\giga\hertz}.
	\begin{figure}
		\centering
		\begin{tikzpicture}%
	\begin{axis}[
		width=.93\linewidth,
		height=.25\textheight,
		xlabel={Frequency Distance $\df$ [\si{\hertz}]},
		ylabel={Received Power $\underline{P_r}$ [\si{\decibel}]},
		xlabel near ticks,
		ylabel near ticks,
		xmin=1e7,
		xmax=1e9,
		xmode=log,
		ymax=-40,
		ymin=-120,
		cycle list name=colorcycle,
		legend pos=south west,
		legend columns=2,%
		legend cell align=left,
		legend style = {
			at = {(1, 1.05)},
			anchor = south east,
			/tikz/every even column/.append style={column sep=0.33cm}
		},
		extra x tick style={grid=none},
		extra y tick style={grid=none},
		ymajorgrids,
		xmajorgrids,
		xminorgrids,
		grid style={line width=.1pt, draw=gray!20},
		major grid style={line width=.25pt,draw=gray!30},
		]
		
		\addplot+[mark repeat=40] table[x=df, y=power] {data/power_envelope_df-1.000000E+08-dist50.0-t10.0-r1.5.dat};
		\addlegendentry{$f_1^{(1)}=\SI{100}{\mega\hertz}$};
		\addplot+[mark repeat=40] table[x=df, y=power] {data/power_envelope_df-2.400000E+09-dist50.0-t10.0-r1.5.dat};
		\addlegendentry{$f_1^{(2)}=\SI{2.4}{\giga\hertz}$};

		\addplot+[densely dashed,ultra thick] coordinates {(5.09761159e+08, -40) (5.09761159e+08, -120)};
		\addlegendentry{Appr./Exact Minimum}
		\addplot+[densely dashed,ultra thick] coordinates {(2.54880579e+08, -40) (2.54880579e+08, -120)};
		\addlegendentry{Appr. Maximum};
		
		\addplot+[dashdotted,ultra thick] coordinates {(1.78373900e+08, -40) (1.78373900e+08, -120) (1.78373900e+08, -130)};
		\addlegendentry{Exact Max. ($f_1^{(1)}$)};
		\addplot+[densely dashdotted,ultra thick] coordinates {(2.53073105e+08, -130) (2.53073105e+08, -120) (2.53073105e+08, -40)};
		\addlegendentry{Exact Max. ($f_1^{(2)}$)};
	\end{axis}
\end{tikzpicture}
		\caption{Received power envelope~$\underline{P_r}$ for two parallel frequencies~$f_1$ and $f_2=f_1+\df$. The additional parameters are set to $d=\SI{50}{\meter}$, $\htx=\SI{10}{\meter}$, and $\hrx=\SI{1.5}{\meter}$. The first peaks are indicated, both the exact locations (numerically determined) and the approximations from \autoref{lem:approx-dw-max-min-sum-power-envelope}. (\autoref{ex:approx-peak-positions})}
		\label{fig:rec-power-df-approximation-peaks}
	\end{figure}
\end{example}

As can be seen from the example above, when using two frequencies~$\omega_1$ and $\omega_2=\omega_1+\dw$, the receive power can be varied for a given distance~$d$ by adjusting $\dw$.
This directly leads to the following optimization of $\domega$.

\subsection{Optimal Frequency Spacing}
Recall from \autoref{prob:opt-prob} that we are interested in the optimization problem
\begin{equation}\label{eq:opt-prob-two-freq}
	\max_{\domega}\min_{d\in[\dmin, \dmax]} \underline{P_{r}}(d, \dw)\,,
\end{equation}
where $\dmin$ and $\dmax$ denote the known interval boundaries of the distance~$d$ between transmitter and receiver.
In order to solve this optimization problem, we need the following characterization.

\begin{lem}\label{lem:reformulation-opt-prob}
	The optimization problem from \eqref{eq:opt-prob-two-freq} can be reformulated as
	\begin{equation}\label{eq:opt-prob-max-min-g}
		\max_{\domega\in[0, \dwmin(\dmax)]} \min \left\{\underline{P_r}(\dmax, \dw), g(\dw)\right\}\,.
	\end{equation}
	with
	\begin{equation}\label{eq:def-func-g}
		g(\dw) = 
		\begin{cases}
			\underline{P_r}(\dmin, \dw) & \text{if}\quad 0 < \dw < \dwmin(\dmin)\\
			\underline{P_r}(d_1, \dw) & \text{if}\quad \dwmin(\dmin) \leq \dw < \dwmin(\dmax)%
		\end{cases}
	\end{equation}
	and
	$\underline{P_r}(d_1, \dw)$ in \eqref{eq:power-d1-two-freq} {at the top of the next page}.
	\begin{figure*}[t]
		\begin{equation}\label{eq:power-d1-two-freq}
			\underline{P_r}(d_1, \dw) = 
			\frac{P_t}{2}\left(\frac{c^2\pi\dw}{2}\right)^2 \!\left(\frac{1}{\omega_1^2}+\frac{1}{\omega_2^2}\right) \!\left(\frac{1}{\sqrt{(c^2\pi^2 - \hrx\htx\dw^2)^2}}-\frac{1}{\sqrt{(c^2\pi^2 + \hrx\htx\dw^2)^2}}\right)^2%
		\end{equation}
		\hrulefill
	\end{figure*}
\end{lem}
\begin{proof}
	The proof can be found in \autoref{app:proof-reformulation-opt-prob}.
\end{proof}

Based on the reformulation of the original optimization problem, we can determine the optimal frequency spacing~$\dw$ for worst-case design as follows.

\begin{thm}[Optimal Frequency Spacing for Worst-Case Design]\label{thm:opt-dw-max-min}
	Consider the described communication system, where two frequencies~$\omega_1$ and $\omega_2=\omega_1+\dw$ are used in parallel.
	The optimal frequency spacing for worst-case design~$\dw^\star$ is given by the intersection of $\underline{P_r}(\dmax, \dw)$ and $g(\dw)$ from~\eqref{eq:def-func-g} in the interval~$\dw\in[\dwmax(\dmin), \dwmax(\dmax)]$, if it exists.
	Otherwise, if no intersection exists, it is given by the maximum of~$\underline{P_r}(\dmax, \dw)$, which is approximately located at
	\begin{equation}\label{eq:dw-opt-approx}
		\dw^\star \approx \frac{c\pi}{\lenref(\dmax)-\len(\dmax)} = \dwmax(\dmax)\,.
	\end{equation}
\end{thm}
\begin{proof}
	The proof can be found in \autoref{app:proof-thm-opt-freq-spacing}.	
\end{proof}

Based on \autoref{thm:opt-dw-max-min}, we can summarize the steps to calculate the optimal frequency spacing~$\df^\star=\w^\star/(2\pi)$ for a worst-case design with given model parameters in \autoref{alg:summary-calculation-opt-freq-spacing}.
The function \texttt{FindIntersection} in Lines~\ref{line:find-intersection-1} and~\ref{line:find-intersection-2} could be any routine that allows calculating the intersection between an increasing and a decreasing function, e.g., by minimizing the (quadratic) distance between them.
A Python implementation with interactive notebooks to reproduce all of the calculations can be found at~\cite{BesserGithub}.
\begin{algorithm*}[t]%
	\caption{Procedure to Find the Optimal Frequency Spacing~$\df^\star$ for Worst-Case Design}
	\label{alg:summary-calculation-opt-freq-spacing}
	\begin{algorithmic}[1]
		\Function{OptimalFrequencySpacing}{$f_1$, $\dmin$, $\dmax$, $\htx$, $\hrx$}
		\State Calculate $\domega_{\pi}(\dmin)$ and $\domega_{\pi}(\dmax)$ \Comment{According to \eqref{eq:dw-maximum}}
		\State Calculate $\dwmin(\dmin)$ and $\dwmin(\dmax)$ \Comment{According to \eqref{eq:dw-minimum}}
		\If{$\underline{P_r}(\dmax, \dw_{\pi}(\dmax) < g(\dw_{\pi}(\dmax))$}\Comment{No intersection}
		\State $\domega^\star=\dw_{\pi}(\dmax)$ \Comment{According to \eqref{eq:dw-opt-approx}}\label{line:simple-case}
		\Else \Comment{An intersection between $\underline{P_r}$ and $g$ exists}
		\If{$\underline{P_r}(\dmax, \widetilde{\dw}(\dmin) > \underline{P_r}(\dmin, \dwmin(\dmin))$} 
		\State \LongComment{Intersection in $[\dw_{\pi}(\dmin), \widetilde{\dw}(\dmin)]$}
		\State $\domega^\star =$ {\Call{FindIntersection}{$\underline{P_r}(\dmin, \dw)$, $\underline{P_r}(\dmax, \dw)$}\label{line:find-intersection-1} \hfill for $\dw_{\pi}(\dmin)<\domega<\dwmin(\dmin)$}
		\Else
		\State \LongComment{Intersection in $[\widetilde{\dw}(\dmin), \dwmin(\dmax)]$}
		\State $\domega^\star =$ {\Call{FindIntersection}{$\underline{P_r}(d_1, \dw)$, $\underline{P_r}(\dmax, \dw)$}\label{line:find-intersection-2} \hfill for $\dwmin(\dmin)<\dw<\dwmin(\dmax)$}
		\EndIf
		\EndIf
		\State $\df^\star=\domega^\star/(2\pi)$
		\State \Return $\df^\star$
		\EndFunction
	\end{algorithmic}
\end{algorithm*}

\begin{rem}[{Complexity and Implementation of \autoref{alg:summary-calculation-opt-freq-spacing}}]\label{rem:complexity}
	The initialization of \autoref{alg:summary-calculation-opt-freq-spacing} (Lines~1--4) and the simple case in Line~\ref{line:simple-case} only require single calculations of \eqref{eq:dw-maximum}, \eqref{eq:dw-minimum}, and the receive power~$\underline{P_r}$.
	This requires basic arithmetic operations, taking roots, and calculating the cosine.
	While the exact complexity of these functions depends on the implementation, all of them are readily available on most modern architectures with highly optimized procedures.
	The computationally expensive part of \autoref{alg:summary-calculation-opt-freq-spacing} is finding the intersection of two functions in Lines~\ref{line:find-intersection-1} and~\ref{line:find-intersection-2}.
	There are several ways of numerically finding the intersection of an increasing and decreasing function in a given interval.
	Two simple approaches are to apply a root-finding method, e.g., Newton's method, to the difference between the two functions, or to minimize the quadratic difference.
	The overall complexity depends on the chosen scheme, its implementation, and the desired accuracy, since a greater accuracy typically requires more iterations, resulting in an increased time complexity.
\end{rem}

\begin{example}[Optimal Frequency Spacing for Worst-Case Design]\label{ex:opt-freq-spacing-sum-power}
As an illustration, we evaluate the following numerical example in detail.
As a base frequency, we choose~$f_1=\SI{2.4}{\giga\hertz}$.
The transmitter is located at height~$\htx=\SI{10}{\meter}$ and the receiver at $\hrx=\SI{1.5}{\meter}$.
While the distribution of the distance between transmitter and receiver is unknown, it is known that the receiver can only be located at a distance between $\dmin=\SI{10}{\meter}$ and $\dmax=\SI{100}{\meter}$.
The optimal frequency gap~$\df$ will therefore lie below~$\widetilde{\df}(\dmax) = \SI{1}{\giga\hertz}$.
In \autoref{fig:opt-prob-min-power-parts}, we show $\underline{P_r}(\dmin, \dw)$, $\underline{P_r}(d_1, \dw)$, and $\underline{P_r}(\dmax, \dw)$ over the frequency spacing~$\df=\dw/(2\pi)$.
First, it can be seen that both $\underline{P_r}(\dmin)$ and $\underline{P_r}(d_1)$ are decreasing.
Recall that these functions define the function~$g$ from~\eqref{eq:def-func-g}.
In contrast, $\underline{P_r}(\dmax)$ increases for $\dw<\dwmax(\dmax)=\SI{502}{\mega\hertz}$.
Next, there exists exactly one intersection of $\underline{P_r}(\dmax)$ and $g$ at around $\df=\SI{177}{\mega\hertz}$.
According to \autoref{thm:opt-dw-max-min}, this corresponds to the optimal frequency spacing~$\df^\star$ that maximizes the worst-case receive power.
The corresponding worst-case power is calculated to $\SI{-85.7}{\decibel}$.
\begin{figure}
	\centering
	\begin{tikzpicture}%
	\begin{axis}[
		width=.92\linewidth,
		height=.26\textheight,
		xlabel={Frequency Spacing $\df$ [\si{\hertz}]},
		ylabel={Received Power $\underline{P_r}$ [\si{\decibel}]},
		xlabel near ticks,
		ylabel near ticks,
		xmax=1E9,
		xmode=log,
		ymax=-60,
		ymin=-140,
		cycle list name=colorcycle,
		extra x ticks={1.76*10^8},
		extra x tick labels={$\df^\star$},
		legend pos=south west,
		legend cell align=left,
		extra x tick style={grid=none},
		extra y tick style={grid=none},
		ymajorgrids,
		xmajorgrids,
		xminorgrids,
		grid style={line width=.1pt, draw=gray!20},
		major grid style={line width=.25pt,draw=gray!30},
		]
		\addplot+[mark repeat=80] table[x=df, y=pmin] {data/power_min_parts1-2.400000E+09-dmin10.0-dmax100.0-t10.0-r1.5.dat};
		\addlegendentry{$\underline{P_r}(\dmin)$};
		
		\addplot+[mark repeat=100] table[x=df, y=pmax] {data/power_min_parts1-2.400000E+09-dmin10.0-dmax100.0-t10.0-r1.5.dat};
		\addlegendentry{$\underline{P_r}(\dmax)$};
		
		\addplot+[mark repeat=40] table[x=df, y=pd1] {data/power_min_parts2-2.400000E+09-dmin10.0-dmax100.0-t10.0-r1.5.dat};
		\addlegendentry{$\underline{P_r}(d_1)$};
		
		\pgfplotsset{cycle list shift=-2}
		\addplot+[mark repeat=40, mark phase=6] table[x=df, y=pmax] {data/power_min_parts2-2.400000E+09-dmin10.0-dmax100.0-t10.0-r1.5.dat};
		
		\addplot[black,dashed,thick] coordinates {(1.76*10^8, -85.7) (1.76*10^8, -150)};
	\end{axis}
\end{tikzpicture}
	\caption{Receive powers~$\underline{P_r}(\dmin)$, $\underline{P_r}(d_1)$, and $\underline{P_r}(\dmax)$ over the frequency spacing~$\df$ for system parameters $f_1=\SI{2.4}{\giga\hertz}$, $\htx=\SI{10}{\meter}$, $\hrx=\SI{1.5}{\meter}$, $\dmin=\SI{10}{\meter}$, and $\dmax=\SI{100}{\meter}$.
	The optimal frequency spacing~$\df^\star$, i.e., the maximum of the minimum of the curves, can be found at around $\df^\star=\SI{177}{\mega\hertz}$. (\autoref{ex:opt-freq-spacing-sum-power})}
	\label{fig:opt-prob-min-power-parts}
\end{figure}

A comparison with the single frequency case is shown in \autoref{fig:opt-df-single-freq-comparison}.
Besides the single frequency case, the actual received power and its lower bound are depicted for the scenario that two frequencies are used in parallel with optimal frequency spacing~$\df=\df^\star$.
When only a single frequency is used, there exist multiple local minima in ${[\dmin, \dmax]}$ with a decreasing receive power.
The global minimum in ${[\dmin, \dmax]}$ for the above parameters can be calculated according to \autoref{thm:min-rec-power-single-freq} to around~\SI{-124.7}{\decibel} at the distance of around \SI{79.4}{\meter}.
In contrast, the received power with two parallel frequencies is always greater than \SI{-85.7}{\decibel}.
Thus, by using a second frequency, the worst-case receive power is improved by around \SI{39}{\decibel}.

\begin{figure}
	\centering
	\begin{tikzpicture}%
	\begin{axis}[
		width=.92\linewidth,
		height=.27\textheight,
		xlabel={Distance $d$ [\si{\meter}]},
		ylabel={Received Power $P_r$ [\si{\decibel}]},
		xlabel near ticks,
		ylabel near ticks,
		xmin=9,
		xmax=110,
		xmode=log,
		ymax=-50,
		ymin=-130,
		cycle list name=colorcycle,
		legend pos=south west,
		legend cell align=left,
		extra x tick style={grid=none},
		extra y tick style={grid=none},
		ymajorgrids,
		xmajorgrids,
		xminorgrids,
		grid style={line width=.1pt, draw=gray!20},
		major grid style={line width=.25pt,draw=gray!30},
		]
		\addplot+[mark repeat=60] table[x=distance, y=powerSingle] {data/power_opt_freq-2.400000E+09-t10.0-r1.5-dmin10.0-dmax100.0.dat};
		\addlegendentry{Single Frequency};
		\addplot+[mark repeat=60] table[x=distance, y=powerOptExact] {data/power_opt_freq-2.400000E+09-t10.0-r1.5-dmin10.0-dmax100.0.dat};
		\addlegendentry{Two Frequencies};
		\addplot+[mark repeat=90] table[x=distance, y=powerOpt] {data/power_opt_freq-2.400000E+09-t10.0-r1.5-dmin10.0-dmax100.0.dat};
		\addlegendentry{Two Freq. -- Lower Bound $\underline{P_r}$};

		\draw[thick, dashed] (10, -85.7) -- (100, -85.7);
		
		\draw[thick, dashed] (10, -124.71) -- (100, -124.71);
		\draw[latex-latex,thick] (85,-85.7) -- node[below,rotate=90] {$\SI{39}{\decibel}$} (85, -124.71);
	\end{axis}
\end{tikzpicture}
	\caption{Received powers for a two-ray ground reflection scenario with parameters $f_1=\SI{2.4}{\giga\hertz}$, $\htx=\SI{10}{\meter}$, $\hrx=\SI{1.5}{\meter}$, $\dmin=\SI{10}{\meter}$, and $\dmax=\SI{100}{\meter}$. First, the receiver power for a single frequency with $f=f_1$ from~\eqref{eq:rec-power-single-freq} is shown. Additionally, the received power for two parallel frequencies with $\df=\df^\star=\SI{177}{\mega\hertz}$ from~\eqref{eq:rec-power-two-freq} and its lower bound from \eqref{eq:rec-power-sum-lower-bound} are depicted. (\autoref{ex:opt-freq-spacing-sum-power})}
	\label{fig:opt-df-single-freq-comparison}
\end{figure}
\end{example}

\begin{rem}[{Optimal Frequency Spacing for Discrete Set of Sub-Carriers}]
	The optimization of the frequency spacing~$\df$ in \autoref{alg:summary-calculation-opt-freq-spacing} is done for continuous values.
	However, in an implementation, e.g., when using \gls{ofdm}, only a discrete set of sub-carriers is available.
	In that case, the optimal frequency spacing~$\df^\star$ can still be calculated using \autoref{alg:summary-calculation-opt-freq-spacing}.
	However, the optimal second frequency~$f_2=f_1+\df^\star$ needs then to be rounded to the closest available frequency from the discrete set of sub-carriers with the best worst-case performance.
	Since the worst-case receive power is continuous with respect to $\df$ and only has a single maximum, this will be the optimal value for the discrete set of frequencies.
\end{rem}

\section{Achievable Rate Analysis}\label{sec:rate-analysis}
After investigating the receive power, we now take a closer look at the resulting achievable data rate.
For a single frequency band of bandwidth~$B$ around the carrier frequency~$\w_1=2\pi f_1$, the capacity is in general given as~\cite[Chap.~5]{Tse2005}
\begin{equation}
	R_1 = \int_{f_1-B/2}^{f_1+B/2} \log_2\left(1 + \frac{P_r(f; P_t/B)}{FN_0}\right) \diff{f}\,,
\end{equation}
where $P_r(f; P_t/B)$ is the receive power at frequency~$f$ and the transmit power is assumed to be equally distributed over the bandwidth~$B$.
Furthermore, $N_0$ is the noise spectral density and $F$ the receiver noise figure.

However, this expression can be simplified for small bandwidths compared to the carrier frequency.
With this narrowband assumption, the capacity can be calculated as~\cite[Chap.~5]{Tse2005}
\begin{equation}\label{eq:rate-single-freq}
	R_1 = B\log_2\left(1 + \frac{P_r(\w_1; P_t)}{BFN_0}\right)\,,
\end{equation}
where $P_r(\w_1; P_t)$ denotes the receive power from~\eqref{eq:rec-power-single-freq} with transmit power~$P_t$.

For a fair comparison, the bandwidth is split into two separate frequency bands of size~$B/2$ in the case of two parallel frequencies.
This yields the total (sum) rate
\begin{multline}
	\label{eq:rate-sum-general}
	R_2 = \frac{B}{2}\log_2\left(1 + \frac{P_{r,1}(\w_1; P_t/2)}{F N_0 \frac{B}{2}}\right) + \\
	\frac{B}{2}\log_2\left(1 + \frac{P_{r,2}(\w_2; P_t/2)}{F N_0 \frac{B}{2}}\right)\,,
\end{multline}
where $P_{r,i}(\w_i)$ again denotes the receive power on frequency~$\w_i$.

As before, it should be emphasized that the transmit power is kept constant.
Hence, it is split in the case of two parallel frequencies.
We again assume an equal power split, i.e., ${\theta=0.5}$.

In \autoref{sec:two-freq}, we optimized the frequency spacing~$\dw$ such that the worst-case (sum) receive power is maximized.
With the following lemma, we show that this result can also be used for worst-case design in terms of the achievable rate.
\begin{lem}\label{lem:rate-freq-spacing}
	Considering the described communication system, where two frequencies~$\w_1$ and~$\w_2=\w_1+\dw$ are used in parallel.
	The lower bound on the achievable rate~$R_2$ is maximized by the optimal frequency spacing~$\dw^\star$ from \autoref{thm:opt-dw-max-min}.
	The worst-case achievable rate~$\underline{R_2}$ in the distance interval ${[\dmin, \dmax]}$ is then given by
	\begin{equation}\label{eq:rate-lower-bound}
		R_2 \geq \underline{R_2} = \frac{B}{2}\log_2\left(1 + \alpha + \frac{\underline{P_r}(d, \dw^\star)}{F N_0 \frac{B}{2}}\right)\,,
	\end{equation}
	where $\underline{P_r}$ is the lower bound on the sum receive power from~\eqref{eq:rec-power-sum-lower-bound} and the offset~$\alpha$ is given by
	\begin{equation}\label{eq:rate-power-offset-alpha}
		\alpha = \left(\frac{1}{F N_0 \frac{B}{2}}\right)^2 \left(\frac{P_t}{2\w_1 \w_2}\right)^2 \left(\frac{c}{2}\right)^4 \left(\frac{1}{\len}-\frac{1}{\lenref}\right)^4\Bigg|_{d=\dmax}\,.
	\end{equation}
\end{lem}
\begin{proof}
	The proof can be found in \autoref{app:proof-lem-rate-freq-spacing}.
\end{proof}

\begin{example}[Achievable Rate Comparison]\label{ex:rate-comparison}
	In order to illustrate the benefit of using two frequencies in terms of achievable rate, we consider the following numerical example.
	We assume the same system parameters as in \autoref{ex:opt-freq-spacing-sum-power}, i.e., $f_1=\SI{2.4}{\giga\hertz}$, $\htx=\SI{10}{\meter}$, $\hrx=\SI{1.5}{\meter}$, $\dmin=\SI{10}{\meter}$, and $\dmax=\SI{100}{\meter}$.
	Based on \autoref{lem:rate-freq-spacing}, the optimal frequency spacing is again $\df^\star=\SI{177}{\mega\hertz}$.
	
	Additionally, we consider a bandwidth of $B=\SI{100}{\kilo\hertz}$.
	Recall that the bandwidth is split in the case of two frequencies, such that each frequency band only has a bandwidth of $\SI{50}{\kilo\hertz}$.
	For the receiver noise, we assume a noise figure of $F=\SI{3}{\decibel}$ and a noise spectral density of $N_0=\SI{-174}{\dBm\per\hertz}$.
	
	In \autoref{fig:rate-comparison}, we show the achievable rates for the single frequency case and the scenario with two frequencies with optimal frequency spacing~$\df=\df^\star$.
	We show both the exact rate~$R_2$ from~\eqref{eq:rate-sum-general} and the lower bound~$\underline{R_2}$ from~\eqref{eq:rate-lower-bound}.
	
	First, it can clearly be seen that the data rate drops significantly at certain distances when only a single frequency is used.
	These distances correspond to the distances~$d_k$ at which the receive power drops to a local minimum, cf. \autoref{sec:single-freq}.
	For the selected system parameters, the lowest rate of around $\SI{51.1}{\kilo\bit\per\second}$ occurs at distance~$d=\SI{79.4}{\meter}$.
	
	In contrast, when employing two frequencies in parallel, the achievable rate is more stable across the distance~$d$.
	The lowest value of the lower bound~$\underline{R_2}$ is evaluated to around $\SI{636.8}{\kilo\bit\per\second}$.
	This amounts to a $12.5\times$~improvement of the worst-case rate compared to the single frequency case.
	Additionally, it should be noted that $\underline{R_2}$ is only a lower bound on the actual rate~$R_2$, which therefore is even higher.
	\begin{figure}[t]
		\centering
		\begin{tikzpicture}%
	\begin{axis}[
		width=.92\linewidth,
		height=.27\textheight,
		xlabel={Distance $d$ [\si{\meter}]},
		ylabel={Data Rate $R$ [\si{\bit\per\second}]},
		xlabel near ticks,
		ylabel near ticks,
		xmin=8,
		xmax=120,
		xmode=log,
		ymin=1e3,
		ymax=1e7,
		ymode=log,
		cycle list name=colorcycle,
		legend pos=south west,
		legend cell align=left,
		legend style = {
			font=\small,
			at = {(.5, .05)},
			anchor = south,
		},
		xtick={10, 100},
		xticklabel style={align=left},
		xticklabels={{$10=\dmin$},{$10^2=\dmax$}},
		extra x tick style={grid=none},
		extra y tick style={grid=none},
		ymajorgrids,
		xmajorgrids,
		xminorgrids,
		grid style={line width=.1pt, draw=gray!20},
		major grid style={line width=.25pt,draw=gray!30},
		]
		\addplot+[mark repeat=60] table[x=distance, y=rateSingle] {data/rate-2.400000E+09-df1.769012E+08-t10.0-r1.5-dmin10.0-dmax100.0-bw1.000000E+05.dat};
		\addlegendentry{Single Frequency $R_1$};
		\addplot+[mark repeat=60] table[x=distance, y=rateTwo] {data/rate-2.400000E+09-df1.769012E+08-t10.0-r1.5-dmin10.0-dmax100.0-bw1.000000E+05.dat};
		\addlegendentry{Two Frequencies $R_2$ ($\df=\df^\star$)};
		\addplot+[mark repeat=90] table[x=distance, y=rateTwoLower] {data/rate-2.400000E+09-df1.769012E+08-t10.0-r1.5-dmin10.0-dmax100.0-bw1.000000E+05.dat};
		\addlegendentry{Two Freq. (Lower Bound) $\underline{R_2}$ ($\df=\df^\star$)};

		\pgfplotsset{cycle list shift=-3}
		\addplot+[very thick, dashed, mark=] coordinates {(10, 51127.3) (100, 51127.3)};
		\pgfplotsset{cycle list shift=4}
		\addplot+[very thick, dashed, mark=] coordinates {(10, 636838.61) (40, 636838.61)};
		
		\draw[latex-latex,thick] (30,51127.3) -- node[left] {$12.5\times$} (30, 636838.61);

		\addplot[black,thick,dashed] coordinates {(10, 1e3) (10, 1e10)};
		\addplot[black,thick,dashed] coordinates {(100, 1e3) (100, 1e10)};
	\end{axis}
\end{tikzpicture}
		\caption{Achievable rates for a two-ray ground reflection scenario with parameters $f_1=\SI{2.4}{\giga\hertz}$, $B=\SI{100}{\kilo\hertz}$, $\htx=\SI{10}{\meter}$, $\hrx=\SI{1.5}{\meter}$, $\dmin=\SI{10}{\meter}$, and $\dmax=\SI{100}{\meter}$.
		First, the rate~$R_1$ for a single frequency with $f=f_1$ from~\eqref{eq:rate-single-freq} is shown.
		Additionally, the achievable rate for two parallel frequencies with $\df=\df^\star=\SI{177}{\mega\hertz}$ from~\eqref{eq:rate-sum-general} and its lower bound from~\eqref{eq:rate-lower-bound} are depicted. (\autoref{ex:rate-comparison})}
		\label{fig:rate-comparison}
	\end{figure}
\end{example}
\section{Outage Probability}\label{sec:outage-prob}
As shown in the previous sections, it can be beneficial to use two frequencies in parallel in order to improve the minimum receive power and achievable data rate.
This directly translates to improving the reliability of the communication system.
Therefore, we analyze the outage probability in the following and show how the proposed scheme enables ultra-reliable communications without perfect \gls{csi} at the transmitter.

Throughout the following, we define that an outage occurs when the achievable rate~$R$ drops below a threshold~$r$~\cite{Tse2005}.
Thus, the outage probability~$\varepsilon$ is given as
\begin{equation}\label{eq:def-outage-prob}
	\varepsilon = \Pr\left(R < r\right)\,.
\end{equation}
When replacing the actual rate~$R$ by the lower bound~$\underline{R}$, we obtain an upper bound~$\overline{\varepsilon}$ on the actual outage probability, i.e., a worst-case bound.

In the following, we assume a random distribution of the distance~$d$ over the interval ${[\dmin, \dmax]}$.
Based on this, we can express the outage probability as
\begin{equation}\label{eq:outage-prob-integral}
	\varepsilon = \int_{R(d) < r} p_{d}(d)\diff{d} %
\end{equation}
where $p_{d}$ denotes the \gls{pdf} of $d$.

Since the integral in \eqref{eq:outage-prob-integral} does not admit a closed-form solution for most common distributions of $d$, we will resort to \gls{mc} simulations in the following.
The source code to reproduce all of the following simulations is available at~\cite{BesserGithub}.

\begin{rem}[Zero-Outage Capacity]\label{rem:zoc}
	Recall that our considered \autoref{prob:opt-prob} is to maximize the minimum receive power, and consequently the minimum rate, over a given interval~${[\dmin, \dmax]}$ of the distance~$d$.
	Thus, all achievable rates in the interval are greater than this minimum.
	Based on this, it can be seen from~\eqref{eq:def-outage-prob} that the outage probability~$\varepsilon$ is zero for rate thresholds~$r$ less than the minimum achievable rate.
	Hence, the minimum achievable rate in ${[\dmin, \dmax]}$ is also the \gls{zoc}~\cite{Besser2021zoc}, which is defined as the maximum rate, such that the outage probability is zero.
	The optimization of the frequency spacing~$\df$ presented in this work, therefore, also maximizes the \gls{zoc} for a given distance interval~${[\dmin, \dmax]}$.
\end{rem}

\begin{example}[Uniform Distribution of the Distance]\label{ex:out-prob-uniform-dist}
	As a first numerical example, we consider a uniform distribution of the distance over the interval~${[\dmin, \dmax]}$.
	In this case, we have $p_{d}(d)=1/(\dmax-\dmin)$.
	While this simplifies the integral in~\eqref{eq:outage-prob-integral}, we are still not able to derive a closed-form solution.
	Thus, the outage probabilities shown in \autoref{fig:out-prob-uniform} are obtained by \gls{mc} simulations with $10^7$~samples.
	The system parameters are again set to the parameters used in \autoref{ex:rate-comparison}. %

	\begin{figure}
		\centering
		\begin{tikzpicture}%
	\begin{axis}[
		width=.93\linewidth,
		height=.27\textheight,
		xlabel={Rate Threshold $r$ [\si{\bit\per\second}]},
		ylabel={Outage Probability $\varepsilon$},
		xlabel near ticks,
		ylabel near ticks,
		xmode=log,
		xmin=1e4,
		xmax=1e7,
		ymax=1,
		ymin=1e-5,
		ymode=log,
		cycle list name=colorcycle,
		legend pos=north west,
		legend cell align=left,
		legend style = {
			font=\small,
			at = {(0.01, .92)},
			anchor = north west,
			fill opacity=0.95,
			text opacity=1,
		},
		extra x tick style={grid=none},
		extra y tick style={grid=none},
		ymajorgrids,
		xmajorgrids,
		xminorgrids,
		grid style={line width=.1pt, draw=gray!20},
		major grid style={line width=.25pt,draw=gray!30},
		]
		\addplot+[mark repeat=40] table[x=threshold, y=singleActual] {data/out_prob_rate-2.400000E+09-dmin10.0-dmax100.0-t10.0-r1.5-bw1.000000E+05.dat};
		\addlegendentry{Single Frequency};
		\addplot+[mark repeat=40] table[x=threshold, y=twoActual] {data/out_prob_rate-2.400000E+09-dmin10.0-dmax100.0-t10.0-r1.5-bw1.000000E+05.dat};
		\addlegendentry{Two Freq. -- Exact};
		\addplot+[mark repeat=40] table[x=threshold, y=twoLower] {data/out_prob_rate-2.400000E+09-dmin10.0-dmax100.0-t10.0-r1.5-bw1.000000E+05.dat};
		\addlegendentry{Two Freq. -- Upper Bound};
				
		\pgfplotsset{cycle list shift=-3}
		\addplot+ coordinates {(51446.48147141248, 1e-5) (51446.48147141248, 3.45e-5)};
		\addplot+ coordinates {(746339.5704254125, 1e-5) (746339.5704254125, 0.0001575)};
		\addplot+ coordinates {(636651.7702164664, 1e-5) (636651.7702164664, 0.0011047)};

		\draw[latex-latex,thick] (51446.48, 2e-5) -- node[above] {$12.4\times$} (636651.77, 2e-5);

	\end{axis}
\end{tikzpicture}
		\caption{%
			Outage probability for a two-ray ground reflection scenario with $\htx=\SI{10}{\meter}$, $\hrx=\SI{1.5}{\meter}$, $f_1=\SI{2.4}{\giga\hertz}$, $\df=\SI{177}{\mega\hertz}$, and $B=\SI{100}{\kilo\hertz}$.
			The distance~$d$ between transmitter and receiver is uniformly distributed between $\dmin=\SI{10}{\meter}$ and $\dmax=\SI{100}{\meter}$, i.e., $d\sim\unif[10, 100]\si{\meter}$.
			The shown outage probabilities are obtained by \gls{mc} simulations with $10^7$~samples. (\autoref{ex:out-prob-uniform-dist})}
		\label{fig:out-prob-uniform}
	\end{figure}
	The first shown outage probability~$\varepsilon$ is for the single frequency case, where $\varepsilon$ is determined by rate~$R_1$ from~\eqref{eq:rate-single-freq}.
	It can be seen that there are zero outages below a rate of around $\SI{51.1}{\kilo\bit\per\second}$, which corresponds to the worst-case rate over the considered distance interval~${[\dmin, \dmax]}$, cf.~\autoref{ex:rate-comparison}.
	As mentioned in \autoref{rem:zoc}, this also corresponds to the \gls{zoc}.
	Above this rate, the outage probability slowly increases.
	In contrast, when using two frequencies in parallel, the increase of the outage probability is much steeper for an increasing rate threshold~$r$.
	This indicates that the rates are more concentrated at similar values over all distances in the considered interval.
	This property can also be observed in \autoref{fig:rate-comparison}.
	It is similar for both the exact outage probability based on $R_2$ from~\eqref{eq:rate-sum-general} and the upper bound~$\overline{\varepsilon}$ determined by $\underline{R_2}$ from~\eqref{eq:rate-lower-bound}.
	Due to this step-like behavior, the $\varepsilon$-outage capacity for small~$\varepsilon$ is significantly higher for two frequencies compared to the single frequency case.
	This includes the \gls{zoc}, which is at around $\SI{636.8}{\kilo\bit\per\second}$ for the upper bound. %
	In the context of ultra-reliable communications, we are typically interested in outage probabilities lower than $10^{-5}$~\cite{Bennis2018}.
	In this regard, the advantage of the proposed frequency diversity scheme can clearly be seen.
\end{example}

However, since the assumption of a uniform distribution of the distance in \autoref{ex:out-prob-uniform-dist} seems arbitrary, we now evaluate a more realistic example of a \gls{uav} flying above flat terrain.
Additionally, we consider a different frequency band in order to demonstrate that our proposed scheme also works for modern 5G~NR frequency bands.

\begin{figure*}
	\centering
	\subfigure[Geometrical setup of the \gls{uav} example.\label{fig:uav-setup}]{%
		\begin{tikzpicture}
	\node[draw,circle,label=below left:{Transmitter},fill=black] (tx) at (0,0) {};
	\node[draw,circle,minimum size=150,align=center,plot1,thick] (lake) at (2.8,2.8) {\hspace*{1em}Lake\\[2em]};
	
	\draw[|<->|,plot0,thick] (lake.center) -- node[below right]{\SI{180}{\meter}} (tx);
	\draw[|<->|,plot2,thick] (lake.center) -- node[below]{\SI{150}{\meter}} (lake.east);
\end{tikzpicture}
	}
	\hfill
	\subfigure[Sample trajectory of the \gls{uav} with indicated outline of the lake.\label{fig:uav-positions}]{%
		\begin{tikzpicture}
	\begin{axis}[
		width=.43\linewidth,
		height=.27\textheight,
		axis equal,
		xlabel={$x$ Position [\si{\meter}]},
		ylabel={$y$ Position [\si{\meter}]},
		ylabel near ticks,
		ylabel shift=-.3em,
		colormap name=viridis,
		xmin=-170,
		xmax=170,
		ymin=-170,
		ymax=170,
		]
		\addplot[scatter,point meta=explicit,line width=2pt,mesh,mark=,] table[x=x,y=y,meta=c] {data/uav_positions.dat};
		
		\draw[plot1,thick,dashed] (axis cs:0,0) circle [radius=150];
	\end{axis}
\end{tikzpicture}
	}
	\caption{Setup of the numerical example of a \gls{uav} flying above flat terrain. The movement of the \gls{uav} is modeled according to~\cite{Smith2022} with movement parameters~$\alpha_1=\alpha_2=1$, $\beta_1=\beta_2=3$, $\gamma_1=\gamma_2=7$, $\sigma_1=\sigma_2=1$, and $s_1=s_2=1$. (\autoref{ex:out-prob-uav})}
\end{figure*}

\begin{example}[\Gls{uav} Flying Above Flat Terrain]\label{ex:out-prob-uav}
	The basic simulation scenario is depicted in \autoref{fig:uav-setup}.
	We consider a flat terrain with a circular lake with a radius of $\SI{150}{\meter}$, which is the area of interest for the \gls{uav} deployment.
	The transmitter is located at a distance of $\SI{30}{\meter}$ from the edge of the lake at height~$\htx=\SI{10}{\meter}$.
	The \gls{uav} flies at height~$\hrx=\SI{3}{\meter}$ above the ground.
	Its movement is modeled according to the mobility model from~\cite{Smith2022}, which is based on \glspl{sde} for each dimension.
	For the \gls{mc} simulation, we fixed the height of the \gls{uav} and only considered the movement in~$x$ and $y$~direction.
	The movement parameters are set to $\alpha_1=\alpha_2=1$, $\beta_1=\beta_2=3$, $\gamma_1=\gamma_2=7$, $\sigma_1=\sigma_2=1$, and $s_1=s_2=1$, cf.~\cite[Sec.~II]{Smith2022}.
	A sample trajectory of the \gls{uav} based on the \gls{sde} model with the described parameters can be found in \autoref{fig:uav-positions}.
	For the simulation of the outage probability, we generate \num{1000}~trajectories with \num{2000}~positions each, i.e., we obtain a total of $2\cdot{}10^6$~samples.
	The source code to reproduce all of the presented results can be found at~\cite{BesserGithub}.

	The base frequency of the first carrier is set to $f_1=\SI{28}{\GHz}$, which lies in the n257~band in the 5G~NR frequency range~2~\cite{etsi5Gfr2}.
	The bandwidth around this carrier is set to $B=\SI{100}{\kHz}$ and the noise parameters are $F=\SI{3}{\decibel}$, and $N_0=\SI{-174}{\dBm\per\hertz}$.
	From the geometrical model, we can derive that the distance~$d$ between transmitter and receiver (\gls{uav}) is between $\dmin=\SI{30}{\meter}$ and $\dmax=\SI{330}{\meter}$.
	Based on \autoref{thm:opt-dw-max-min}, the optimal frequency spacing for this setup is $\df=\df^\star=\SI{191}{\mega\hertz}$.
	Since the 5G~NR standard allows for total channel bandwidths of more than $\SI{200}{\MHz}$~\cite[Sec.~5.3]{etsi5Gfr2}, both carrier frequencies~$f_1$ and $f_2=f_1+\df^\star$ could lie within the same 5G~channel.
	\begin{figure}
		\centering
		\begin{tikzpicture}%
	\begin{axis}[
		width=.92\linewidth,
		height=.27\textheight,
		xlabel={Rate Threshold $r$ [\si{\bit\per\second}]},
		ylabel={Outage Probability $\varepsilon$},
		xlabel near ticks,
		ylabel near ticks,
		xmode=log,
		xmin=1e1,
		xmax=1e7,
		ymax=1,
		ymin=1e-7,
		ymode=log,
		cycle list name=colorcycle,
		legend pos=north west,
		legend cell align=left,
		legend style={%
			font=\small,
			fill opacity=.95,
			text opacity=1,
			anchor=west,
			at={(.02, .52)},
		},
		extra x tick style={grid=none},
		extra y tick style={grid=none},
		ymajorgrids,
		xmajorgrids,
		xminorgrids,
		grid style={line width=.1pt, draw=gray!20},
		major grid style={line width=.25pt,draw=gray!30},
		]
		\addplot+[mark repeat=40] table[x=threshold, y=singleActual] {data/out_prob_uav-2.800000E+10-dmin30.0-dmax330.0-t10.0-r3.0-bw1.000000E+05-df1.900763E+08.dat};
		\addlegendentry{Single Frequency};
		\addplot+[mark repeat=40] table[x=threshold, y=twoActual] {data/out_prob_uav-2.800000E+10-dmin30.0-dmax330.0-t10.0-r3.0-bw1.000000E+05-df1.900763E+08.dat};
		\addlegendentry{Two Freq. -- Exact};
		\addplot+[mark repeat=40] table[x=threshold, y=twoLower] {data/out_prob_uav-2.800000E+10-dmin30.0-dmax330.0-t10.0-r3.0-bw1.000000E+05-df1.900763E+08.dat};
		\addlegendentry{Two Freq. -- Upper Bound};
		\addplot+[mark repeat=40] table[x=threshold, y=twoComparison] {data/out_prob_uav-2.800000E+10-dmin30.0-dmax330.0-t10.0-r3.0-bw1.000000E+05-df1.900763E+08.dat};
		\addlegendentry{Two Freq. -- $\df=\SI{100}{\MHz}$};
		
		\pgfplotsset{cycle list shift=-3}
		\addplot+ coordinates {(109652.87172540072, 1e-7) (109652.87172540072, 5.4e-05)};
		\addplot+ coordinates {(83168.7128036327, 1e-7) (83168.7128036327, 1.79e-05)};
		\addplot+ coordinates {(34337.4376583475, 1e-7) (34337.4376583475, 0.00036)};

		\draw[latex-latex,thick] (17.75, 1e-5) -- node[below,fill=white,draw=none,] {$4686\times$} (83168.7, 1e-5);
	\end{axis}
\end{tikzpicture}
		\caption{%
			Outage probability for a two-ray ground reflection scenario of a \gls{uav} flying above flat terrain.
			The system parameters are $\htx=\SI{10}{\meter}$, $\hrx=\SI{3}{\meter}$, $f_1=\SI{28}{\giga\hertz}$, $\df^\star=\SI{191}{\mega\hertz}$, and $B=\SI{100}{\kilo\hertz}$.
			Additionally, the upper bound for $\df=\SI{100}{\MHz}$ is shown for comparison.
			The distance~$d$ between transmitter and receiver is randomly distributed between $\dmin=\SI{30}{\meter}$ and $\dmax=\SI{330}{\meter}$.
			The shown outage probabilities are obtained by \gls{mc} simulations with $2\cdot{}10^6$~samples. (\autoref{ex:out-prob-uav})}%
		\label{fig:out-prob-uav}
	\end{figure}
	
	The resulting outage probabilities for the single frequency and two frequencies scenario are shown in \autoref{fig:out-prob-uav}.
	Similarly to \autoref{ex:out-prob-uniform-dist}, it can be seen that the outage probability for a single frequency decreases slowly when decreasing the rate threshold~$r$.
	In contrast, the achievable rates are more concentrated when using two frequencies in parallel, which results in a steeper outage probability curve in \autoref{fig:out-prob-uav}.
	This results in a higher $\varepsilon$-outage capacity for small~$\varepsilon$.
	Assuming that the application can tolerate an outage probability of up to $\varepsilon=10^{-5}$, the rate would have to be adjusted to be less than around $\SI{18}{\bit\per\second}$ for a single frequency.
	In contrast, when using two frequencies in parallel with the optimal frequency spacing~$\df=\df^\star$, the rate could be up to $\SI{83.2}{\kilo\bit\per\second}$, while still fulfilling the same reliability constraint.
	This corresponds to a gain of nearly $4700\times$ for the $\varepsilon$-outage capacity with $\varepsilon=10^{-5}$.
	For comparison, we additionally show the upper bound on the outage probability for the non-optimized frequency spacing~$\df=\SI{100}{\MHz}$ in \autoref{fig:out-prob-uav}.
	While the outage performance is still better than with only a single frequency, the benefit of choosing the optimized frequency spacing is clearly visible.

	This practical example highlights the significant reliability improvements that are possible when using two frequencies in parallel with the optimal frequency spacing.
	It should also be emphasized that this scheme only requires knowledge of the interval of possible distances between transmitter and receiver \emph{without} requiring perfect \gls{csi} at the transmitter.
\end{example}
\section{Conclusion}\label{sec:conclusion}
In this work, we have investigated the receive power, achievable rate, and outage probability in two-ray ground reflection scenarios with unknown distance between transmitter and receiver.
We have shown that using two frequencies in parallel can significantly improve the reliability over using only a single frequency.
In particular, we derived the optimal frequency spacing such that the worst-case receive power is maximized.
An especially useful aspect of the presented results is that only very limited knowledge is required at the transmitter.

All of the results have been evaluated for various numerical examples, including a realistic scenario of a \gls{uav} flying above flat terrain.
This allows a quantitative classification of the performance improvement by the proposed scheme.

While only two parallel frequencies are considered in this work, it could be extended to diversity systems with multiple frequencies in future work.
This is a promising research direction to further improve the reliability of the described communication systems.
Additionally, when adding the assumption that the transmitter has knowledge of the distribution of the distances between transmitter and receiver, the optimization problem could be modified to minimize the outage probability for the given distribution.
An initial study of this problem can be found in~\cite{Besser2023asilomar}.
Furthermore, it will be an interesting aspect in future work to drop the assumption of ideal \gls{mrc} and consider implementation issues of combining the signals on the two frequencies.

\appendices
\section{Proof of \autoref{prop:worst-case-rho}}\label{app:proof-prop-worst-case-params}
First, we start with the general receive power from~\eqref{eq:rec-power-single-freq-params}, which we rewrite as
\begin{equation}
	P_r = a \left(\frac{1}{\len^2} + b^2 \frac{1}{\lenref^2} + 2b \frac{1}{\len\lenref}\cos(\Delta\phi)\right)
\end{equation}
using the shorthand notation~$b=\rho\reflcoeff$ and the constant~$a=P_t\left(\frac{c}{2\w}\right)^2$.
Based on the discussion of the parameters in \autoref{sec:model-problem-formulation}, the range of~$b$ is determined as $-1 < b < 0$.
Thus, the destructive interference occurs for $\cos\Delta\phi = 1$, which gives the corresponding receive powers as $a \left(\frac{1}{\len^2} + b^2 \frac{1}{\lenref^2} + 2b \frac{1}{\len\lenref}\right)$.
The worst-case~$b$ can be found by setting the derivative to zero,
\begin{equation*}
	\frac{2}{\lenref} \left(\frac{b}{\lenref} + \frac{1}{\len}\right) = 0 \quad \Rightarrow \quad b = \rho\reflcoeff = -\frac{\lenref}{\len} < -1\,,
\end{equation*}
where the last inequality directly follows from the geometrical model.
Additionally, it follows that the receive power at positions of destructive interference decreases with decreasing~$b$.
Therefore, the infimum for the feasible range of~$b$ is given when setting $b=\rho\reflcoeff = -1$.
\section{Proof of \autoref{lem:rec-power-lower-bound-two-freq}}\label{app:proof-rec-power-lower-bound}
The lower bound on $P_r$ from~\eqref{eq:rec-power-two-freq} is calculated as the (lower) envelope of the function, which is given as the absolute value of its analytic function~\cite{Bracewell2000}.
However, we are only interested in bounding the oscillating part of~$P_r$ given by the cosine terms as
\begin{equation*}
	s = \frac{\theta\cos\left(\frac{\omega_1}{c}(\lenref-\len)\right)}{\omega_1^2} + \frac{\bar{\theta}\cos\left(\frac{\omega_2}{c}(\lenref-\len)\right)}{\omega_2^2}\,.
\end{equation*}
The analytic function of~$s$ is given as $s+\imag \hat{s}$, where $\hat{s}=\hilbert\{s\}$ is the Hilbert transform~$\hilbert$ of $s$.
The envelope of~$s$ is then given as $\abs{s+\imag\hat{s}}$.
With the correspondence~$\hilbert\{\cos(\omega t)\}=\sin(\omega t)$~\cite{King2009hilbert}, we obtain the analytic signal
\begin{multline*}
	s + \imag\hat{s} = \\
	\frac{\theta\cos\left({\omega_1 t}\right)}{\omega_1^2} + \frac{\bar{\theta}\cos\left({\omega_2}t\right)}{\omega_2^2} + \imag\left(\frac{\theta\sin\left({\omega_1}t\right)}{\omega_1^2} + \frac{\bar{\theta}\sin\left({\omega_2}t\right)}{\omega_2^2}\right),
\end{multline*}
where we use the shorthand~$t=\frac{\lenref-\len}{c}$.
The absolute value can then be calculated as
\begin{equation*}
	\abs{s+\imag\hat{s}}^2 = \frac{\theta^2}{\omega_1^4} + \frac{\bar{\theta}^2}{\omega_2^4} + \frac{2\theta\bar{\theta} \cos\left((\omega_2 - \omega_1)t\right)}{\omega_1^2 \omega_2^2}\,.
\end{equation*}
Applying the definitions of $t$ and $\dw=\omega_2-\omega_1$ and substituting the envelope of $s$ into~\eqref{eq:rec-power-two-freq}, we obtain~\eqref{eq:rec-power-sum-lower-bound}.
\section{Proof of \autoref{lem:reformulation-opt-prob}}\label{app:proof-reformulation-opt-prob}
Similarly to the minimum received power from~\eqref{eq:min-rec-power-single-freq} in the single frequency case, the minimum received power in the two frequency case is given as the minimum over the boundary points and the lowest local minimum peak at distance~$d_k$.
As before, the worst case among the $d_k$ is found at $d_1$.
Therefore, if ${d_1\in[\dmin, \dmax]}$, the minimum of~$\underline{P_r}$ is given as
\begin{multline*}
	\min_{d\in[\dmin, \dmax]} \underline{P_{r}}(d, \dw) = \\
	\min \left\{\underline{P_r}(\dmin, \dw), \underline{P_r}(\dmax, \dw), {\underline{P_r}(d_1, \dw)}\right\},
\end{multline*}
for a given value of $\dw$.

However, since we can adjust the frequency gap~$\domega$, we can influence the distance~$d_1$ at which the drop in received power occurs.
In order to achieve this drop at a given distance~$d_1$, the frequency spacing~$\dw$ needs to be adjusted as $\dw=\dwmin(d_1)$ based on~\eqref{eq:dw-minimum}. %
With this, we obtain $\underline{P_r}(d_1, \dw)$ according to~\eqref{eq:power-d1-two-freq}.
As mentioned above, in the worst case, we have that ${d_1\in[\dmin, \dmax]}$.
In order for this to happen, $\domega$ needs to be within
\begin{equation}
	\dwmin(\dmin) \leq \domega < \dwmin(\dmax)\,.
\end{equation}
In this case, it is straightforward to verify that $\underline{P_r}(d_1, \dw) < \underline{P_r}(\dmin, \dw)$, hence, the minimum receive power is determined as the minimum between $\underline{P_r}(d_1, \dw)$ and $\underline{P_r}(\dmax, \dw)$.

For $0 < \domega < \dwmin(\dmin)$, there is no local minimum in ${[\dmin, \dmax]}$ and the minimum receive power is given as the minimum between $\underline{P_r}(\dmin)$ and $\underline{P_r}(\dmax)$.
Therefore, we introduce the auxiliary function~$g$ as
\begin{equation*}
	g(\dw) = 
	\begin{cases}
		\underline{P_r}(\dmin, \dw) & 0 < \dw < \dwmin(\dmin)\\
		\underline{P_r}(d_1, \dw) & \dwmin(\dmin) \leq \dw < \dwmin(\dmax)\,.
	\end{cases}%
\end{equation*}
Combining all of the above, the optimization problem from~\eqref{eq:opt-prob-two-freq} can then be rewritten as~\eqref{eq:opt-prob-max-min-g}.
\section{Proof of \autoref{thm:opt-dw-max-min}}\label{app:proof-thm-opt-freq-spacing}
From \autoref{lem:reformulation-opt-prob}, we know that the optimal frequency spacing~$\dw^\star$ is given as the solution to~\eqref{eq:opt-prob-max-min-g}.
In order to solve~\eqref{eq:opt-prob-max-min-g}, we start with the simple observation that $\underline{P_r}$ approaches a finite positive value for $\dw=0$, for which additionally $\underline{P_r}(\dmin, 0) > \underline{P_r}(\dmax, 0)$ holds since $\dmin < \dmax$.

First, we prove the second part of the theorem, when no intersection between~$g$ and~$\underline{P_r}(\dmax)$ exists.
Since $\underline{P_r}(\dmax) < g$ holds for $\dw=0$ and no intersection exists, we have that $\underline{P_r}(\dmax) < g$ for the whole domain $0\leq\dw<\dwmin(\dmax)$.
Hence, the minimum between $\underline{P_r}(\dmax)$ and $g$ is simply $\underline{P_r}(\dmax)$, and the optimization problem~\eqref{eq:opt-prob-max-min-g} reduces to determining the maximum of $\underline{P_r}(\dmax)$.
By \autoref{lem:approx-dw-max-min-sum-power-envelope}, this can approximately be found at~$\dw^\star=\dw_{\pi}$, i.e.,~\eqref{eq:dw-opt-approx} from the statement of the theorem.

Next, we consider the case that an intersection between $\underline{P_r}(\dmax)$ and $g$ exists.
It is straightforward to verify that $\underline{P_r}(d_1)$ is strictly decreasing for $\dwmin(\dmin)\leq \dw < \dwmin(\dmax)$.
Based on \autoref{lem:approx-dw-max-min-sum-power-envelope}, we have established that $\underline{P_r}(\dmin)$ is increasing for $\dw < \dw_{\pi}(\dmin)$ and decreasing for $\dw_{\pi}(\dmin)<\dw<\dwmin(\dmin)$.
Hence, $g$ is increasing for $\dw < \dw_{\pi}(\dmin)$ and decreasing for $\dw_{\pi}(\dmin) < \dw < \dwmin(\dmax)$.
Similarly, we find that $\underline{P_r}(\dmax)$ is increasing up to $\dw_{\pi}(\dmax)$ and decreasing on the rest of the domain.

From the definitions of $\len$ and $\lenref$, it can easily be seen that their difference~$\lenref-\len$ decreases for an increasing distance~$d$.
Thus, it follows that $\dw_{\pi}(\dmax) > \dw_{\pi}(\dmin)$, i.e., the maximum of $\underline{P_r}(\dmax)$ occurs at a higher frequency distance~$\dw$ than the maximum of~$\underline{P_r}(\dmin)$.
Additionally, it follows from $\dmax>\dmin$ that $\underline{P_r}(\dmin, \dw_{\pi}(\dmin)) > \underline{P_r}(\dmax, \dw_{\pi}(\dmin))$ and $\underline{P_r}(\dmin, \dw_{\pi}(\dmin)) > \underline{P_r}(\dmax, \dw_{\pi}(\dmax))$, due to the additional path loss between~$\dmin$ and~$\dmax$.
Combining all of the above observations, we can conclude that one intersection of $\underline{P_r}(\dmax)$ and $g$ occurs at ${\dw^\star\in[\dw_{\pi}(\dmin), \dw_{\pi}(\dmax)]}$ since $\underline{P_r}(\dmax)$ is strictly increasing and $g$ strictly decreasing in this interval.
For $\dw<\dw^\star$, it follows that $\min\{\underline{P_r}(\dmax), g\}$ is increasing, while it is decreasing for $\dw>\dw^\star$.
Hence, the maximum of $\min\{\underline{P_r}(\dmax), g\}$ occurs at the intersection~$\dw^\star$.
\section{Proof of \autoref{lem:rate-freq-spacing}}\label{app:proof-lem-rate-freq-spacing}
The rate~$R_2$ from~\eqref{eq:rate-sum-general} can be reformulated as
\begin{multline*}
	R_2 =\\
	\frac{B}{2} \log_2\left(1 + \frac{P_{r,1}(\w_1) + P_{r,2}(\w_2)}{F N_0 \frac{B}{2}} + \frac{P_{r,1}(\w_1) P_{r,2}(\w_2)}{\left(F N_0 \frac{B}{2}\right)^2}\right).
\end{multline*}
The product of the individual receive powers~$P_{r,i}(\w_i)$ from~\eqref{eq:rec-power-single-freq} can be lower bounded by
\begin{align*}
	P_{r,1}(\w_1) P_{r,2}(\w_2)
	&\geq \left(\frac{P_t}{2\w_1 \w_2}\right)^2 \left(\frac{c}{2}\right)^4 \left(\frac{1}{\len}-\frac{1}{\lenref}\right)^4\\
	&\overset{(a)}{\geq} \left(\frac{P_t}{2\w_1 \w_2}\right)^2 \left(\frac{c}{2}\right)^4 \left(\frac{1}{\len}-\frac{1}{\lenref}\right)^4\Bigg|_{d=\dmax}\,,
\end{align*}
where $(a)$ follows from the fact that $\frac{1}{\len}-\frac{1}{\lenref}$ is a decreasing function in the distance~$d$, thus, the lowest value is achieved at the maximum distance~$\dmax$.
Combining this with the noise power leads to the offset~$\alpha$ from~\eqref{eq:rate-power-offset-alpha}
\begin{equation*}
	\alpha = \left(\frac{1}{F N_0 \frac{B}{2}}\right)^2 \left(\frac{P_t}{2\w_1 \w_2}\right)^2 \left(\frac{c}{2}\right)^4 \left(\frac{1}{\len}-\frac{1}{\lenref}\right)^4\Bigg|_{d=\dmax}\,.
\end{equation*}
With the monotonicity of the logarithm, this already establishes the relation
\begin{equation*}
	R_2 \geq \frac{B}{2} \log_2\left(1 + \alpha + \frac{P_{r,1}(\w_1) + P_{r,2}(\w_2)}{F N_0 \frac{B}{2}}\right)\,.
\end{equation*}
From this, it also follows directly that
\begin{multline*}
	\min_{d\in[\dmin, \dmax]} R_2 \geq \\
	\qquad\frac{B}{2} \log_2\bigg(1 + \alpha + \frac{2}{F N_0 B} \min_{d\in[\dmin, \dmax]} P_{r,1}(\w_1) + P_{r,2}(\w_2)\bigg). \!\!\!\!\!
\end{multline*}
Due to the monotonicity of the logarithm, the minimum rate is maximized when the argument is maximized, i.e.,
\begin{multline*}
	\max_{\domega}\min_{d\in[\dmin, \dmax]} R_2 \geq \\
	\frac{B}{2} \log_2\bigg(1 + \alpha + \frac{2}{F N_0 B} \max_{\domega}\min_{d\in[\dmin, \dmax]} P_{r,1}(\w_1) + P_{r,2}(\w_2)\bigg). \!\!\!\!\!
\end{multline*}
The solution to the inner problem on the right-hand side is given by \autoref{thm:opt-dw-max-min}.
Hence, we can also apply it for worst-case design of the sum rate~$R_2$.

The minimum sum receive power by \autoref{thm:opt-dw-max-min} is given as~$\underline{P_r}(d, \dw^\star)$, which in turn leads to the lower bound on the achievable rate
\begin{equation*}
	R_2 \geq \underline{R_2} = \frac{B}{2}\log_2\left(1 + \alpha + \frac{\underline{P_r}(d, \dw^\star)}{F N_0 \frac{B}{2}}\right)\,,
\end{equation*}
which is~\eqref{eq:rate-lower-bound} and concludes the proof.

\printbibliography

\end{document}